\documentclass[aps,twocolumn,floatfix,tightenlines,prb,superscriptaddress]{revtex4-2}
\usepackage{graphics}
\usepackage{subcaption}
\usepackage{epstopdf}
\usepackage{epsfig}
\usepackage{graphicx}
\usepackage{epsf,epic}
\usepackage{color}
\usepackage{amsmath}
\usepackage{booktabs}
\usepackage{multirow}
\usepackage{physics}
\usepackage{hyperref}
\usepackage{amsfonts}
\usepackage{wrapfig}
\usepackage{breqn}
\usepackage{pstricks}
\usepackage[utf8]{inputenc}
\usepackage{multirow}
\usepackage{fancyref}
\usepackage{pst-node}
\usepackage{soul}
\usepackage{float}
\usepackage{bm}
\usepackage{dcolumn}
\newcommand{\etal}{\textit{et al.\ }}
\newcommand{\ie}{\textit{i.e.\ }}

\makeatletter
\usepackage{etoolbox} 
\appto{\appendix}{%
	\@ifstar{\def\theequation@prefix{A.}}%
	{}%
}
\preto\maketitle{%
  \begingroup\lccode`~=`,
  \lowercase{\endgroup
  \let\saved@breqn@active@comma~
  \let~}\active@comma 
}
\appto\maketitle{%
  \begingroup\lccode`~=`,
  \lowercase{\endgroup
  \let~}\saved@breqn@active@comma 
}
\makeatother
\begin{document}
\title{Charge disproportionation and spin ordering in LiV$_2$O$_5$ and NaV$_2$O$_5$: a distributed moment point of view}
\author{Claudio Garcia}
\affiliation{Department of Physics, Case Western Reserve University, 10900 Euclid Avenue, Cleveland, Ohio 44106-7079, USA}
\affiliation{National Laboratory of the Rockies, Golden, Colorado 80401, USA}
\author{Ilteris K. Turan}
\affiliation{Department of Physics, Case Western Reserve University, 10900 Euclid Avenue, Cleveland, Ohio 44106-7079, USA}
  \author{Jerome Jackson}
  \affiliation{Scientific Computing Department, STCF Daresbury Laboratory, Warrington WA4 4AD, United Kingdom}
  \author{Swagata Acharya}
  \affiliation{National Laboratory of the Rockies, Golden, Colorado 80401, USA}
  \author{Mark van Schilfgaarde}
  \affiliation{National Laboratory of the Rockies, Golden, Colorado 80401, USA}
  \author{Walter R L. Lambrecht}
\email{walter.lambrecht@case.edu}
\affiliation{Department of Physics, Case Western Reserve University, 10900 Euclid Avenue, Cleveland, Ohio 44106-7079, USA}
\begin{abstract}

  The effects of symmetry breaking charge disproportionation between V$^{4+}$ and V$^{5+}$ on the exchange interactions  and related magnetic, electronic and optical properties are studied in LiV$_2$O$_5$ and  NaV$_2$O$_5$. The magnetic exchange interactions between local moments on atomic sites are calculated using a linear response approach based on the quasiparticle self-consistent QS$GW$ band structure. This approach differs from the usual localized spin model in which a single electron spin is centered on the V-O$_b$-V rung molecular orbital model.
The spin is indeed found to be distributed over the two V and while the induced magnetic moments are small on the vanadyl O$_v$ above the V  and bridge O$_b$ linking them, their exchange interactions with the V are found to be of the same order of magnitude as between V atoms. Besides the antiferromagnetic coupling between nearest neighbors along the {\bf b}-direction of the zigzag chains considered as only interaction in most previous literature, several additional exchange interactions are found to be important: the ferromagnetic coupling between V across a bridge O$_b$, and  the antiferromagnetic interactions between the vanadyl O$_v$ and bridge O$_b$ with their nearest neighbor V. The lower  magnetic moments on the V$^{5+}$ than the V$^{4+}$ in the $P2_1mn$ structure lead to more but not completely isolated chains. The exchange interactions are used to calculate the N\'eel temperature $T_N$ for antiferromagnetic ordering within the mean-field (MFA) and random phase (RPA) approximations for two different crystal structures with $P2_1mn$ and $Pmmn$ spacegroups, where the first one includes charge disproportionation and the second does not. All critical temperatures, calculated in the MFA from a classical Heisenberg model with unit vector spins, were found to be significantly lower than the experimental  $T_N$. This indicates the importance of treating the spins quantum mechanically by including a $(S+1)/S$ correction factor which for $S=1/2$ amounts  to a factor 3 correction.  In the $P2_1mn$ structure, the exchange interactions involving the small moment V$^{5+}$ atoms are reduced while those involving the large moment V$^{4+}$ are increased. This leads to a larger $T_N$ in the MFA but not in the RPA where the 1D character of the model produces a stronger reduction of $T_N$. For NaV$_2$O$_5$, the RPA for the $Pmmn$ structure gives good agreement with the experimental ordering temperature identified with the temperature of the maximum in the susceptibility of 320 K. For LiV$_2$O$_5$ a similar value is obtained but no experimental value is available. 
 Band structures, optical dielectric functions calculated using  the Bethe Salpeter Equation method and spin waves are also calculated and presented for both crystal structures and both compounds.  The spin wave dispersions are calculated from the classical Heisenberg Hamiltonian including magnetic moments on both V and O atoms. Besides the lowest energy weakly interacting spin waves of the two chains of spins in the unit cell, the model predicts additional high energy optical spin waves, which correspond to more complex spin excitations in which the spins on the two  V across a bridge, vanadyl O and bridge O precess  independently from each other instead of staying locked together according to their collinear alignment in the ground state. These show only weak dispersion and extend up to about 1 eV but we argue that these are an artefact of using the Heisenberg model beyond its validity range  and would be suppressed in a more complete theory that allows for coupling of the transverse spin with longitudinal spin and charge excitations.  
\end{abstract}
\maketitle
\section{Introduction}
The V$_2$O$_5$ bronzes obtained by intercalating layered  V$_2$O$_5$ with Li or Na
form an interesting family of materials, which has attracted attention in the context of battery cathodes\cite{Whittingham1976}, electrochromic applications \cite{Jarry_2020} and  catalysis \cite{Reshma2024}. Furthermore, the unique band structure of  V$_2$O$_5$ which exhibits a split-off lowest narrow conduction band with almost one-dimensional dispersion also leads to magnetic properties when it becomes partially filled. Essentially, the half-filling of this narrow band leads to a Mott-insulating behavior.
NaV$_2$O$_5$ attracted particular attention when it was first proposed to undergo a spin Peierls transition \cite{Isobe1996} at 34 K.

\begin{figure}[h]
\begin{subfigure}{8cm}\centering
  \includegraphics[width=8cm]{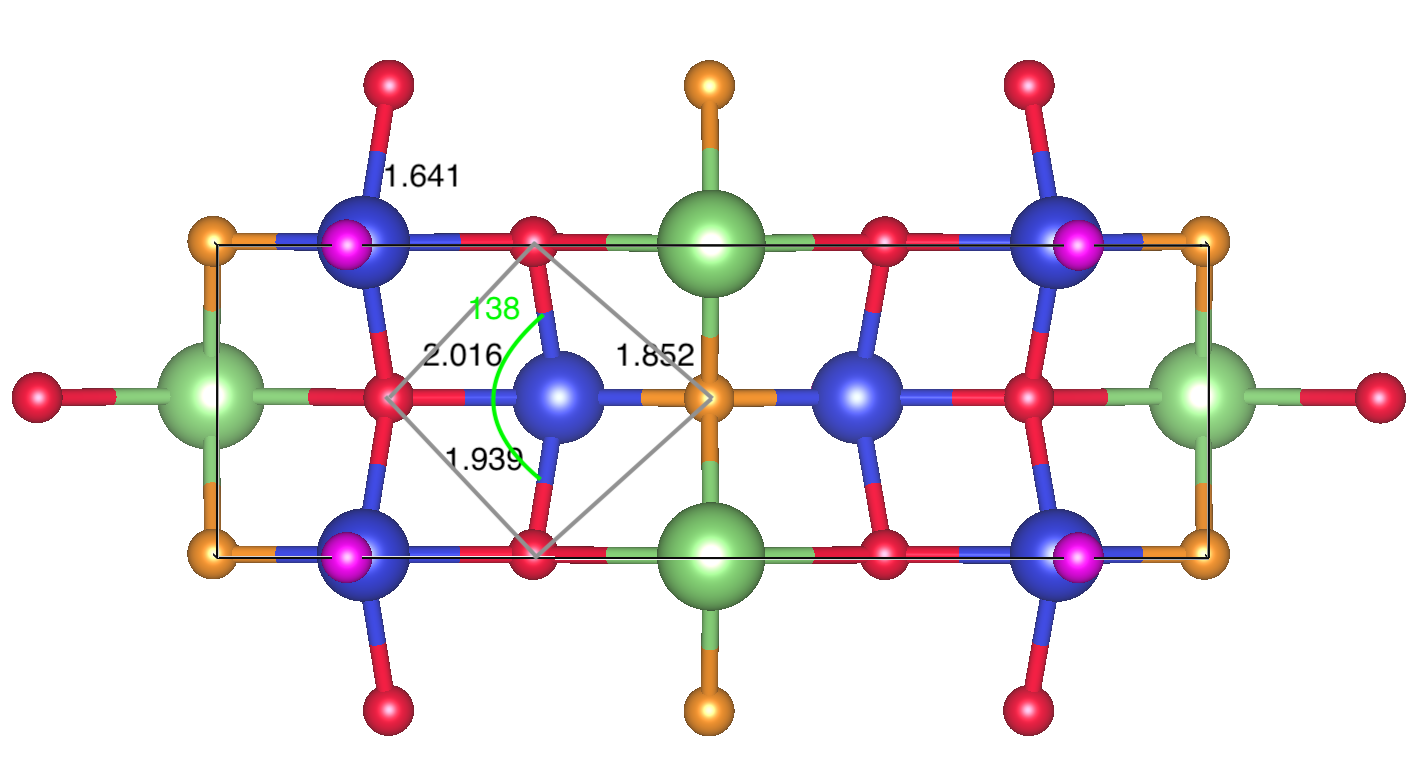}{}
\caption{}
\end{subfigure}
\begin{subfigure}{8cm}\centering
  \includegraphics[width=8cm]{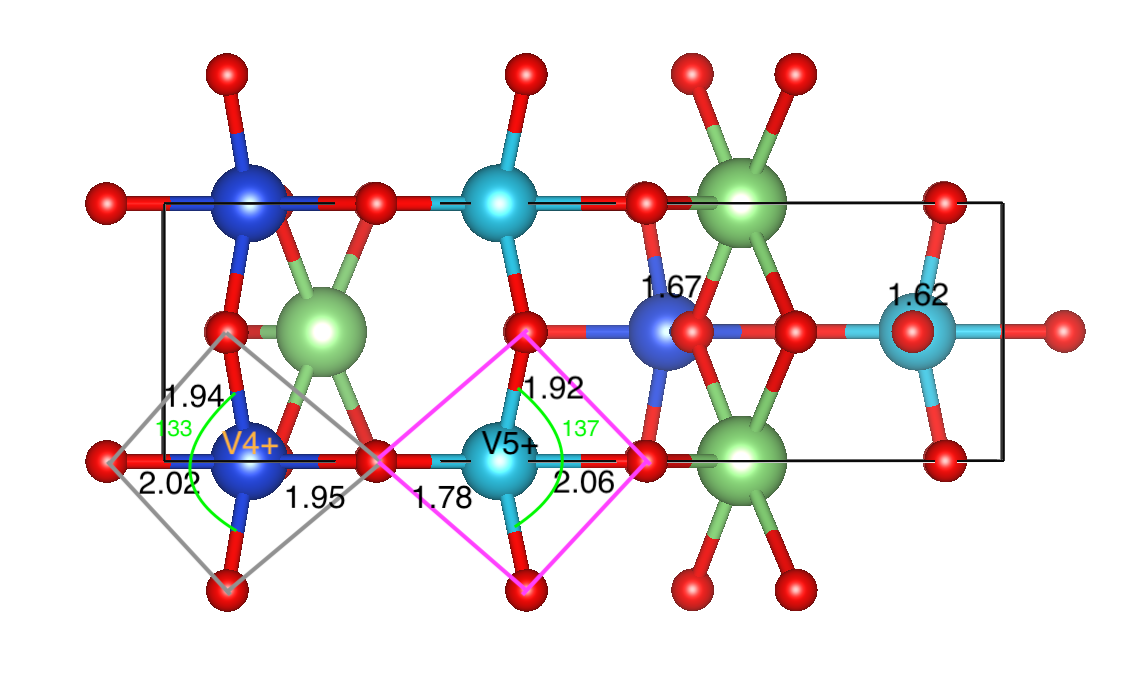}{}
  \caption{}
\end{subfigure}
  \caption{Crystal structure of  LiV$_2$O$_5$ in (a) the $Pmmn$ structure and (b) the $P2_1mn$ structure;  V atoms blue, Li green, bridge O orange, vanadyl O, magenta and chain O red.
    The bond distances around V are indicated in \AA, and the bond angles along the zigzag chain in degrees, In (b) the V$^{4+}$ are dark blue and the V$^{5+}$  light blue and all O are just colored red. The unit cell is also shifted from case (a) but this has no consequence.  This figure was made with VESTA3 \cite{Vesta}.\label{figstructures}}
\end{figure}
The crystal structure of V$_2$O$_5$ contains double zigzag chains of V
along the crystallographic {\bf b}-direction in the layers (perpendicular to ${\bf c}$)  connected by bridge oxygens in the ${\bf a }$ direction,
and with so-called vanadyl-oxygen atoms pointing alternating up and down from the V perpendicular to the layer. These vanadyl O are essentially bonded to a single V because of the large distance to the V in the next layer, which is only weakly van der Waals bonded. In LiV$_2$O$_5$, Li is situated in between the layers and in the large interstital site or channel as shown in Fig. \ref{figstructures}. 
In Na- or LiV$_2$O$_5$, there is one electron per V$_2$O$_5$ unit and the split-off band being half filled, a magnetic moment forms by splitting the band in spin up and spin down bands\cite{Popovic1999,Bhandari_2015_Na}. However, the question now is whether this electron is shared equally between two V across a bridge or localizes on one of the two. These two possibilities are shown in (a) and (b) in Fig. \ref{figstructures}, where we have indicated the structural differences in terms of bond lengths and angles. In the symmetry preserving  $Pmmn$ structure, the V are nominally $V^{+4.5}$. In the symmetry broken $P2_1mn$ structure, one can see that the O atoms have moved away from the V$^{4+}$ where the electron is localized and the Li has moved closer to it. This distortion may be viewed as a self-trapped polaron.  
In the $P2_1mn$ case we have an isolated 1D chain of V$^{4+}$ with one spin $S=1/2$ per atom  separated from the next chain by an inert chain of V$^{5+}$ atoms.
This one dimensional chain of spins turns out to have antiferromagnetic (AFM) Heisenberg type exchange couplings, essentially because of the superexchange via the almost 180$^\circ$ indirect coupling in V-O$_c$-V bonds. This so-called 1D nearest neighbor coupled AFM Heisenberg chain was predicted \cite{Pytte74} to have a spin-singlet ground state in which nearest neighbor pairs of spins form a singlet state and hence, when including interaction with the phonons,  dimerize the chain forming a spin gap in the spin wave spectrum in direct analogy to the opening of a band gap in a half-filled 1D band known as the Peierls transition. This so-called {\sl charge disproportionated} structure of isolated chains was at first thought to be the crystal structure of NaV$_2$O$_5$ and differs from the V$_2$O$_5$ space group by the absence of a mirror-plane perpendicular to the {\bf a}-axis and passing through the bridge oxygens. The space group is then $P2_1mn$ instead of $Pmmn$. The electronic structure in the $P2_1mn$ structure was studied in the LDA+U approximation  by Popovi\'c and Vukajlovi\'c\cite{Popovic1999} and found to have a gap of 2.7 eV and a magnetic moment of 0.97 $\mu_B$ and 0.05 $\mu_B$ on the other V atom, while  only a very small gap was obtained at the LDA  level.  

At high temperature an  isolated AFM Heisenberg chain was found  to exhibit a broad maximum in its susceptibility rather than a well defined N\'eel critical temperature because the fluctuations in 1D prevent the formation of long-range order. The temperature of this broad maximum, $T_{max}$ was predicted by simulations on finite chains\cite{BonnerFisher64} to occur at $T_{max}=1.282|J|/k$ where $J<0$ is the nearest neighbor exchange coupling and in NaV$_2$O$_5$ is found at about 320$\pm50$ K. Actually, this is not so far from a simple mean field estimate $T_c=\frac{2}{3}|2J|/k$.  Assuming this is the only relevant exchange interaction leads to an estimate of $J\approx-21$ meV. It was first measured by Carpy \etal \cite{Carpy72} in a study of NaV$_2$O$_{5-x}$F$_x$ oxyfluorides. In contrast, the oxyfluorides with $x\ne0$ show a more distinct sharp phase transition and a signficantly lower $T_N$ of about 140-155 K. Interestingly, the fluorides were assumed to exhibit the higher symmetry $Pmmn$ phase with all equal V$^{4.5+}$ ions while the pure oxide was assumed to have the lower symmetry charge disproportionated $P2_1mn$ phase. 

The same broad maximum around 320 K was observed by Isobe and Ueda\cite{Isobe1996} and used by them to extract a $J\approx-280$ K $\approx -24$ meV between the V-localized spins on the isolated chain but in addition they found a sharp drop in susceptibility at 34 K and consistent with the known structure of  NaV$_2$O$_5$ at that time assigned it to the interesting quantum spin-Peierls transition. A spin gap of 9.8 meV was confirmed by inelastic neutron scattering by Fujii\etal \cite{Fujii97} but already  indicated a more complex unit cell.

However, in 1998  \cite{VonSchnering98,Smolinski1998}, a new crystal structure refinement showed that at room temperature the structure did exhibit mirror planes of the $Pmmn$ space group and thus has  V$^{4.5+}$.
While at first puzzling how to reconcile this with an insulating state, it was then concluded  that the single spin resides in a molecular type state on the V-O$_b$-V unit, spread equally over the two V. This compound was then identified as  a quarter filled ladder compound with V-O$_b$-V rungs. This invalidated the spin-Peierls interpretation and generated considerable interest \cite{Horsch1998,Ueda2012,Konstantinovic2001,Mostovoy1999,Cuoco1999,Presura2000,Long1999}. Eventually this was settled by assigning the  34 K transition to be a charge disproportionation transition  and this was
confirmed experimentally by NMR spectroscopy\cite{Ohama1999} and by using anomalous X-ray diffraction \cite{Nakao2000} in which X-rays  resonant with the V $K$-absorption edge were used to provide sensitivity of the X-ray scattering to the charge state allowing to distinguish the  V$^{4+}$ from  V$^{5+}$. However, it was further shown that below the transition, the ordering is not along 1D chains but of a zigzag type involving a larger $2\times2\times4$ unit cell\cite{Fujii97,Ludecke1999,deBoer2000}.
Recently, Korotin\cite{Korotin2025} showed that the zigzag ordering of V$^{4+}$ and V$^{5+}$ has lower total energy than the chain ordering using a constrained LDA+U approach for the related NH$_4$V$_2$O$_5$. The charge disproportionation was studied using a cluster dynamical mean field theory (DMFT) by Mazurenko \etal\cite{Mazurenko2002} using a minimal cluster of the 
V-O$_b$-V rung and including not only on-site but also inter-site repulsive  Coulomb interactions. They concluded that dynamical charge fluctuations play a crucial role above the 34 K critical temperature. Similar conclusions were reached  using a $t-J-V$ model in \cite{Horsch1998,Cuoco1999}.

While the DMFT study of Mazurenko \etal \cite{Mazurenko2002} consider the charge disproportionation
as resulting from inter-site Coulomb interactions, it  appears also to be related to polaronic distortion. It is well known that adding electrons to pure V$_2$O$_5$, self-trapped polarons can form by locally distorting the structure, \ie moving the nearby O away from the V where the electron is trapped \cite{Scanlon08,Watthaisong_2019,Ngamwongwan_2021}. One can clearly see this local distortion in the $P2_1mn$ structural model compared to the $Pmnn$ structure in Fig. \ref{figstructures}. While the $P2_1mn$ model is now understood not to be the actual crystal structure of Na(Li)V$_2$O$_5$, it still provides us with a reasonable model that exhibits charge disproportionation. We use it here to study the behavior of the exchange interactions and to what extent a model of isolated chains can be validated. Furthermore, we will study whether these exchange interactions provide a reasonable estimate of the $T_N$.

While the bulk of the previous works have focused on the intriguing low temperature 
transition, the spin ordering temperature of $T_{max}$ has received less attention. 
Recently, several works have tried to model spin fuctuations not with DMFT but rather using spatial fluctuation models. In the present material  it appears that both charge fluctuations and spin fluctuations need to be taken into account.
In this paper our goal is not to describe the low temperature charge disproportionated state  but rather to understand the nature of the magnetic exchange interactions from first principles and study  how they depend on whether charge disproportionation is present or not.

In contrast to the prevailing point of view, in which the magnetic properties
are interpreted in terms of a single localized spin per V-O$_b$-V rung, we take a {\em distributed moment} point of view. It is clear
from the previous work that the spin density is spread over the two V and the O in
between and possibly even beyond it to the vanadyl oxygens. 
In the present work  we associate magnetic moments with each atomic sphere and calculate their exchange interactions. In some sense, this represents the average over time of a single spin fluctuating between sites. In spin-dependent  density functional theory it is natural to think in terms of a spin density and here we coarse-grain this description by lumping this spin density into magnetic moments per atomic sphere. 
To calculate the exchange interactions  between these moments we use  a linear response approach in which the classical Heisenberg type exchange interactions are obtained without {\sl a-priori} assumptions on the range of the interactions. The magnetic moments are thus viewed as classical vectors with a direction that can point in any direction and the size of the magnetic moments is incorporated in the exchange interactions by writing the Heisenberg Hamiltonian as interactions between unit vectors on each site, 
$H=-\sum_{ij}J_{ij}{\bf e}_i\cdot{\bf e}_j$. This approach can be justified based on adiabatic decoupling of the longitudinal and transverse degrees of freedom of the magnetic moments 
within a rigid spin approximation (RSA)  of the spin density in each atomic sphere\cite{Antropov1996}. The RSA here means that the shape of the spin density
in each sphere stays fixed and is collinar within a sphere so that a well-defined moment with fixed magnitude exists in each sphere. This is less restrictive than assuming that the overall spin $S=1/2$ is rigid. The validity of this approximation will be discussed later in the paper when we discuss the spin waves. 
They are extracted using the approach proposed by Kotani  and van Schilfgaarde\cite{Kotani2008} by calculating  a site averaged transverse spin susceptibility  $D^\pm_{aa'}({\bf q},\omega)$ from which Heisenberg real space exchange interactions are extracted by taking the static limit, $\omega=0$ of the inverse of this susceptibility matrix
and applying a  Fourier transform to real space,
$J_{ij}({\bf q})=\sum_{{\bf T}}e^{i{\bf q}\cdot{\bf T}} J_{ij}^{\bf T}$. The transverse spin susceptibility is calculated from the energy band structure eigenvalues and eigenvectors and these are here obtained within the QS$GW$ method.

We caution that the mapping on a spin $S$ Heisenberg type model using $M=2S$ where $M$ is the local magnetic moment may lead to a somewhat problematic interpretation  when the local moment is about 0.5 $\mu_B$ and corresponds to a $S=1/4$.
Even smaller moments are induced on oxygen atoms and are found to play a key role. 
We should keep in mind that in the end this represents a single electron with $S=1/2$ which happens to be spread over two V atoms and the O in between  in its wavefunction or can be thought of as fluctuating between these sites. To take into account the quantum mechanical nature of the spins in the critical temperatures, we will find it necessary to take into account a $(S+1)/S$ correction factor. 

Most of the investigations of the magnetic and charge disproportionation aspects have focused on NaV$_2$O$_5$ because Li$_x$V$_2$O$_5$ at Li concentration between 1 and 2 is commonly found in the somewhat more complex $\gamma$-structure. Similar issues in that structure were studied by Valent\'\i\ \etal \cite{Valenti2001}. Because of their similarity, we here consider both Li and NaV$_2$O$_5$ in the layered structure.

\section{Computational Methods}
The band structure approach used in this work is the full-potential linearized muffin-tin orbital method FP-LMTO as implemented in the {\sc Questaal} codes \cite{questaalpaper}.
We start from the density functional theory (DFT) with PBEsol  \cite{PBEsol} as exchange-correlation functional and add Hubbard $U$ to ensure an initial magnetic moment. We then calculate the first iteration in the $GW$ approximation of Hedin's many-body-perturbation theory (MBPT) \cite{Hedin65,Hedin69} and from there on switch off the Hubbard $U$ and let the $GW$ method converge within the QS$GW$ scheme\cite{Kotani07,Ismail-Beigi_2017}.
The $GW$ self-energy thus takes over the role of $U$. The quasiparticle self-consistency means that from the frequency dependent self-energy $\Sigma(\omega)=iG(\omega)\otimes W(\omega)$ where $\otimes$ is a convolution, $G$ is  the one-electron Green's function and $W$ the screened Coulomb interaction, we extract an energy independent non-local exchange-correlation potential,
$\tilde\Sigma=\frac{1}{2}|\psi_i\rangle {\rm Re}\left[\Sigma_{ij}(\epsilon_i)+\Sigma_{ij}(\epsilon_j)\right]\langle\psi_j|$, which replaces the initial DFT+U exchange-correlation potential, which is then made self-consistent. Here ${\rm Re}$ means taking the Hermitian part and summation over repeated indices is assumed.  The real part of the quasiparticle  eigenvalues then become the same as the non-interacting Kohn-Sham  eigenvalues and  in that sense the method is quasiparticle self-consistent and becomes independent of the DFT starting point.

To calculate exchange interactions, we use the linear response approach introduced by Kotani and van Schilfgaarde  \cite{Kotani2008}. This approach starts from the expression for the bare transverse spin susceptibility,
\begin{equation}
		\begin{split}
			\chi_{\bm{q}}^{0+-}(\bm{r}, \bm{r}',\omega)&=\\\sum_{\bm{k}n\downarrow}^{occ.}\sum_{\bm{k}'n'\uparrow}^{unocc.} \frac{\psi^{*}_{\bm{k}n\downarrow}(\bm{r}) \psi_{\bm{k}'n'\uparrow}(\bm{r})\psi^{*}_{\bm{k}'n'\uparrow}(\bm{r}')\psi_{\bm{k}n\downarrow}(\bm{r}')}{\omega-(\epsilon_{\bm{k}'n'\uparrow}-\epsilon_{\bm{k}n\downarrow})+i\delta}\\
			+\sum_{\bm{k}n\downarrow}^{unocc.}\sum_{\bm{k}'n'\uparrow}^{occ.}\frac{\psi^{*}_{\bm{k}n\downarrow}(\bm{r}) \psi_{\bm{k}'n'\uparrow}(\bm{r})\psi^{*}_{\bm{k}'n'\uparrow}(\bm{r}')\psi_{\bm{k}n\downarrow}(\bm{r}')}{-\omega-(\epsilon_{\bm{k}n\downarrow}-\epsilon_{\bm{k}'n'\uparrow})+i\delta},\label{eq:chi}
		\end{split}	
	\end{equation} 
where $\bm{k}'=\bm{k}+\bm{q}$, and $\psi_{{\bf k}n\sigma}$ and $\epsilon_{{\bf k}n\sigma}$ are the spin-dependent band structure eigenstates and eigenvalues, which in our case are calculated in the QS$GW$ approximation. This spin-susceptibility is then integrated over muffin-tin spheres  within the RSA and converted to a site labeled $D^{0+-}_{ab}({\bf q},\omega)$. This involves a projector $\bar{e}_a({\bf r})$ in the spheres constructed from the local magnetization density $m_a({\bf r})=n_{a\uparrow}({\bf r})-n_{a\downarrow}({\bf r})$ by dividing by the total moment $M_a$, $e_a({\bf r})=m_a({\bf r})/M_a$ and then renormalizing
$\bar{e}_a({\bf r})=e_a({\bf r})/N_a$ whith $N_a=\int_ad^3r e^2_a({\bf r})$. In practice, $\chi_{IJ}^{0+-}({\bf q},\omega)$ is first constructed in  the basis of the $GW$ mixed-product basis functions $B^{\bf q}_I({\bf r})$ and one then uses the overlap of these with the projector $\bar{e}_a({\bf r})$ to obtain
\begin{equation}
  D_{ab}^{0+-}({\bf q},\omega)=\langle\bar{e}_a|B^{\bf q}_I\rangle \chi_{IJ}^{0+-}({\bf q},\omega) \langle B^{\bf q}_J|\bar{e}_b\rangle,\\
\end{equation}

The reason for this way of coarse-graining to a sphere and somewhat unusual normalization of the site projectors $\bar{e}_a({\bf r})$ was not elaborated on in the paper that introduced it\cite{Kotani2008}. One can think of it as follows. The $[\chi^{0+-}_{\bf q}]^{-1}({\bf r},{\bf r}',\omega)$ describes the interaction between spin densities $m({\bf r})$. In order to obtain interactions between total moments per sphere, we need to integrate over the spheres surrounding the two sites $a,b$  with a weight function proportional to the spin density.
Assuming for simplicity  that $m_a({\bf r})=(M_a/V_a)\theta_a({\bf r})$, in other words that it is a constant density in the sphere, then $e_a({\bf r})=\theta_a({\bf r})/V_a$ and integrates to 1, but $\bar{e}_a({\bf r})=\theta_a({\bf r})$.
By using $\bar{e}_a$ we actually integrate over the sphere with a step function as projector, while using $e_a$ we would have divided by volume and dimensionally still obtained an interaction between densities instead of between total moments. More realistically, if we assume the spin density has a Gaussian form  with width $\sigma\ll s_a$, \ie  well localized inside the sphere, so that we can approximate radial integrals over the sphere by an integral up to infinity, then $e_a(r)=\frac{1}{\pi^{3/2}\sigma^3}\exp{-r^2/\sigma^2}$ and $\bar{e}_a(r)=\frac{1}{2^{3/2}}\exp{-r^2/\sigma^2}$. The un-normalized $e_a(r)$ would then be inversely proportional to the volume over which the spin density extends, while normalizing to $\bar{e}_a$ removes the volume factor. While other projectors could in principle be considered, the present scheme has the advantage that the integration over a sphere is done weighted by a projector with the shape of the spin density  in the sphere. Its validity as a scheme to obtain effective inter-site exchange interactions is justified in practice by the fact that it yields excellent agreement with spin wave energies obtained from other methods, such as Liechtenstein's well known Green's function multiple scattering theory based linear response method \cite{Liechtenstein1987}, the recent time dependent (TD)DFT approach \cite{Binci2025}, and other methods based on linear response \cite{Savrasov2006} as tested on well-known materials like Fe, MnO, NiO.
With the exception of Liechtenstein's method, the spin waves in these appraoches are directly obtained from the
poles of the susceptibility without extracting site exchange interactions but these can then be fitted to the dispersion curves. 
When used with LDA band structures, it tends to overestimate spin wave energies but as shown in \cite{Kotani2008} it yields excellent agreement with experiment for spin waves when used with QS$GW$ band structures. 

We can also easily justify why the inverse of $\chi^{0+-}$ is needed for the exchange interactions. Essentially, $m({\bf r})=\delta E/\delta B({\bf r})$ where $E$ is the total energy and $B({\bf r})$ is the magnetic field, or in the context of collinear spins, the difference between up and down spin potentials, and $\chi^{0+-}({\bf r},{\bf r}')=\delta m({\bf r})/\delta B({\bf r}')=\delta^2E/\delta B({\bf r})\delta B({\bf r}')$.
Meanwhile, from the Heisenberg Hamiltonian as written in terms of interactions between magnetic moments, it is clear that $J({\bf r},{\bf r}')=\delta^2E/\delta m({\bf r})\delta m({\bf r}')$. Converting from
fields to magnetization densities as basic variables, it is clear that $J$ is the inverse of $\chi$. Of course we here should view this in the context of
density functional theory as the response to a periodic perturbation in the spin density of wavevector ${\bf q}$ and time dependent with frequency $\omega$.

In \cite{Kotani2008} the dynamic dressed spin-susceptibility is obtained from
\begin{eqnarray}
          [D({\bf q},\omega)]^{-1}_{aa'}&=&[D^0({\bf q},\omega)]^{-1}_{aa'}+\frac{\omega}{M_a}\delta_{aa'}\nonumber \\
          &&-M_a^{-1}\sum_bM_b[D^0({\bf q}=0,\omega)]^{-1}_{ba}\delta_{aa'},\label{eq:DQ}
\end{eqnarray}
under the assumption that the interaction kernel $U({\bf q},\omega)$ is ${\bf q}$-independent and on-site only and so that the susceptibility satisfies the $\omega\rightarrow0$ and $\omega\rightarrow\infty$ sumrules. This inverse of the susceptibility is then shown to be equivalent to  the matrix of exchange interactions 
\begin{equation}
  \bar{J}_{ab}({\bf q})=J_{ab}({\bf q})-\left(\sum_c\frac{1}{S_a}J_{ac}(0)S_c\right)\delta_{ab}
\end{equation}
that describes spin-wave excitations for right and left circular spin-waves, 
\begin{equation}
  \sum_b\left(\frac{\omega}{2S_a}\delta_{ab}\mp \bar{J}_{ab}({\bf q})\right) S^{\pm}_b({\bf q})=0, \label{eq:spwv}
\end{equation}
of a Heisenberg Hamiltonian
$H^{\cal H}=-\sum_{a,{\bf T}}\sum_{b,{\bf T}'} J_{a{\bf T},b{\bf T'}} {\bf S}_{a{\bf T}}\cdot {\bf S}_{b{\bf T}'}$ if we make the static approximation, $\omega=0$ in $[D({\bf q},\omega)]^{-1}$ and assign $M_a=2S_a$. Note that here the Heisenberg exchange interactions are written with respect to the assumed spins on each site and all pairs are counted twice. We can convert this to a classical Heisenberg Hamiltonian
written
\begin{equation}
  H^{cH}=-\sum_{aa'} J^c_{aa'}{\bf e}_a\cdot{\bf e}_{a'}\label{jclas}
\end{equation}
  in terms of unit vectors by incorporating the magnitude of the spin (or magnetic moment) in the 
  $J^c_{aa'}=|S_a|J_{aa'}|S_{a'}|$.

  Note that this approach makes no {\sl a-priori} assumptions about the range of the exchange interactions but also may lead to spins  that are not restricted to integer or half-integer values. These really just represent half the magnetic moments per site.

  From, the real space exchange interactions we can then calculate the critical temperature using the mean field or other approximations.
    For a system with several magnetic sites per unit cell,  \cite{Sasioglu2004} the mean field $T_c$ is obtained by diagonalizing the matrix of the $J^0_{ab}=\sum_{\bf T} J_{ab}^{0{\bf T}}$, where if
  $a=b$, the on-site term $J_{aa}^{00}$ is excluded from the sum and we here use the classical Heisenberg interactions defined as above  with respect to unit vectors, 
\begin{equation}
  \sum_b[J_{ab}^0-j\delta_{ab}]\langle S_b\rangle=0.\label{eq:tc}
\end{equation}
  The mean-field critical temperature is then given by $k_BT_c=2j_{max}/3$ with $j_{max}$
  the highest eigenvalue and $\langle S_b\rangle$ represents the mean field average spin on sublattice $b$. For a quantum mechanical spin, the equation should be multiplied by a factor $(S+1)/S$, which is particularly important for small $S$. However, because quantum mechanical spins not equal to a half-integer or integer make no sense, at this point we will return to the  intepretation of the spin as the  spin of a single electron but with a wavefunction spread over different sites. In other words we will use $S=1/2$ at this point in the correction factor, which then becomes a factor 3. 

 We  also use the random phase approximation (RPA) of Tyablikov and Callen \cite{Tyablikov1959,Callen63,Rusz2005}. In this method, one constructs the matrix $\textbf{N}$, with elements
        \begin{equation}
        N_{ab}({\bf q})=-S_a\bar{J}_{ab}({\bf q}),
        \label{eqNmatrix}
        \end{equation} 
whose eigenvalues give the spin waves and in which $S_a$ is now the thermal average of the spins on sublattice $a$.
The $J_{ab}=J^c_{ab}/|S_aS_b|$ are now exchange interactions
of the Hamiltonian written in terms of spins $S_a$.
      One then calculates the Brillouin zone average of the 
        inverse of the $N$-matrix, and uses  the consistency equation \cite{Rusz2005}
        \begin{equation}
          \langle S_a \rangle=\frac{2}{3}\beta_c\left\{\int_{BZ} d^3q [N^{-1}({\bf q};\langle S_a \rangle)]_{aa}\right\}^{-1}, \label{eqIa}
        \end{equation}
        to determine $\beta_c=1/k_BT_c$.
        In practice the thermal average spin vectors at $T_c$ become scale independent and are replaced by a normalized vector and one obtains at each iteration 
        \begin{equation}
			T_c^{RPA}=\frac{2}{3/k_B}\frac{1}{I_a\langle S_a\rangle},
        \end{equation}
        where $I_a$ is the Brillouin zone integral in Eq.\ref{eqIa}
        calculated in terms of the $\langle S_a \rangle_{\text{old}}$ initiated by the normalized eigenvector of the mean field ${\bf J}({\bf 0})$ matrix and iterated until the normalized  $\langle S_a \rangle$ no longer changes, \ie at the fixed point of the consistency equation. We check that each sub-lattice gives the same $T_c$ and that the final $\langle S_a\rangle$ are close to the initial one from the MFA and correspond to the spin arrangement in the reference configuration. 
        
        The difficulty is that $\textbf{N}({\bf q}=0)$ diverges. To avoid the divergence one calculates instead the Green's function $G(z)=[z{\bf I}+{\bf N}({\bf q}=0)]^{-1}$ for some points $z$ in the complex plane  with positive imaginary part and uses analytic continuation towards the real axis at $z\rightarrow0$.\cite{Rusz2005,Hass1984} The distance of the points in the complex plane is set by a parameter $\epsilon$ and the adequacy of the approximation is guided by the requirement that the imaginary part of the inverse is small compared to the real part. The RPA method requires a fine ${\bf q}$-mesh to adequately sample the region near ${\bf q}\rightarrow 0$.
       
  \section{Results}
  The structural parameters used in this paper were obtained from experiment or from Materials Project (MP) \cite{MPcitation} and are detailed in
  Supplemental Information (SI)\cite{supinfo}. We note that for Li$_x$V$_2$O$_5$, several different structures occur as the Li content increases. In  the $\gamma$-structure, the layers are significantly distorted with the vanadyl oxygens on either side of V-O$_b$-V rung pointing in opposite directions. Here, we do not consider this structure but only the layered structure that stays close to the V$_2$O$_5$ structure and is known as the $\alpha^\prime$ structure for NaV$_2$O$_5$ and as the $\delta$-phase for LiV$_2$O$_5$.  The phase diagram and detailed structural information for  Li$_x$V$_2$O$_5$ can be found in
  \cite{Rocquefelte2003} and \cite{Delmas1994}. 
 \subsection {Total energies}
 The difference in total energies between the $Pmmn$ and $P2_1mn$ and between the AFM and FM ordering are studied using  the GGA+U  approach.
 \begin{table}
   \caption{Total energy differences  in eV per formula unit.\label{totE}}
   \begin{ruledtabular}
\begin{tabular}{ccc}
  \multicolumn{3}{c}{LiV$_2$O$_5$}\\ \hline
  FM-AFM&$Pmmn$   & 0.017\\
  FM-AFM&$P2_1mn$ & 0.015 \\
  $Pmmn-P2_1mn$ & AFM & 0.820 \\
  & relaxed AFM & 1.096 \\
  $Pmmn-P2_1mn$ & FM & 0.822\\ 
     \end{tabular}
   \end{ruledtabular}
\end{table}
For LiV$_2$O$_2$ we find that in both structures the AFM state has lower energy than the FM state by a  similar amount of about 15 meV/formula unit.
We also find the symmetry broken $P2_1mn$ structure to have lower energy than the symmetric $Pmmn$ structure by about 0.82 eV /formula unit in the structures provided by MP. This amounts to 0.1 eV per atom, while the MP lists a difference in energy of formation of 0.034  eV but also finds the $P2_1mn$ to have the lower energy.  When we relax the AFM structure, the energy difference further increases. We also find that the amount of charge disproportionation decreases somewhat in the relaxed structure.  However, it is not {\sl a-priori} clear that the relaxation of the structure based on the present DFT+U method is preferable over the relaxation with the functional used in MP.
Therefore we base the rest of the results on the structure obtained from MP as it brings out more clearly the effects of the charge disproportionation. 
\begin{figure*}
  \begin{subfigure}{8cm}\centering
   \includegraphics[width=8cm]{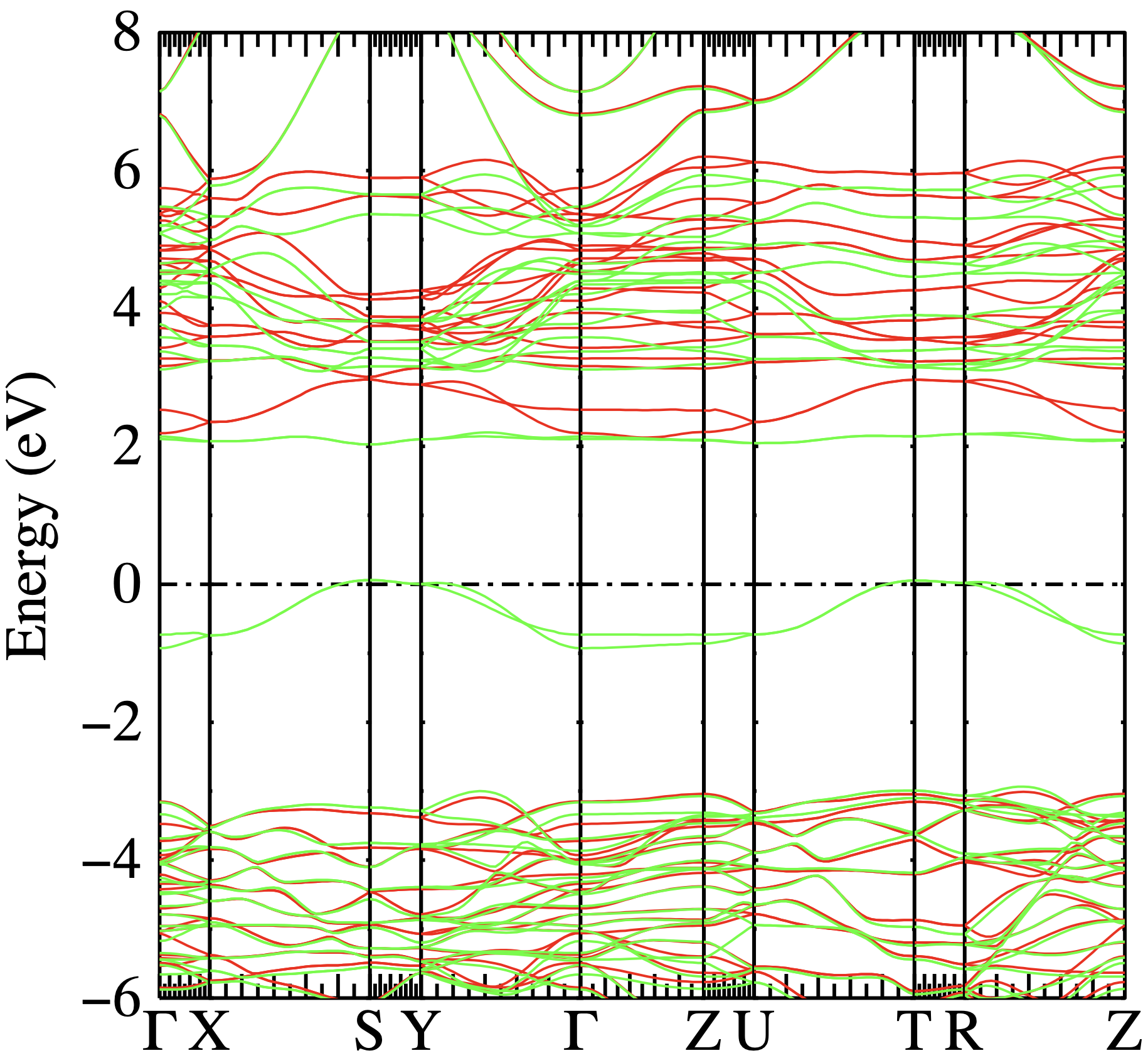}{}
   \caption{}
  \end{subfigure}
    \begin{subfigure}{8cm}\centering
      \includegraphics[width=8cm]{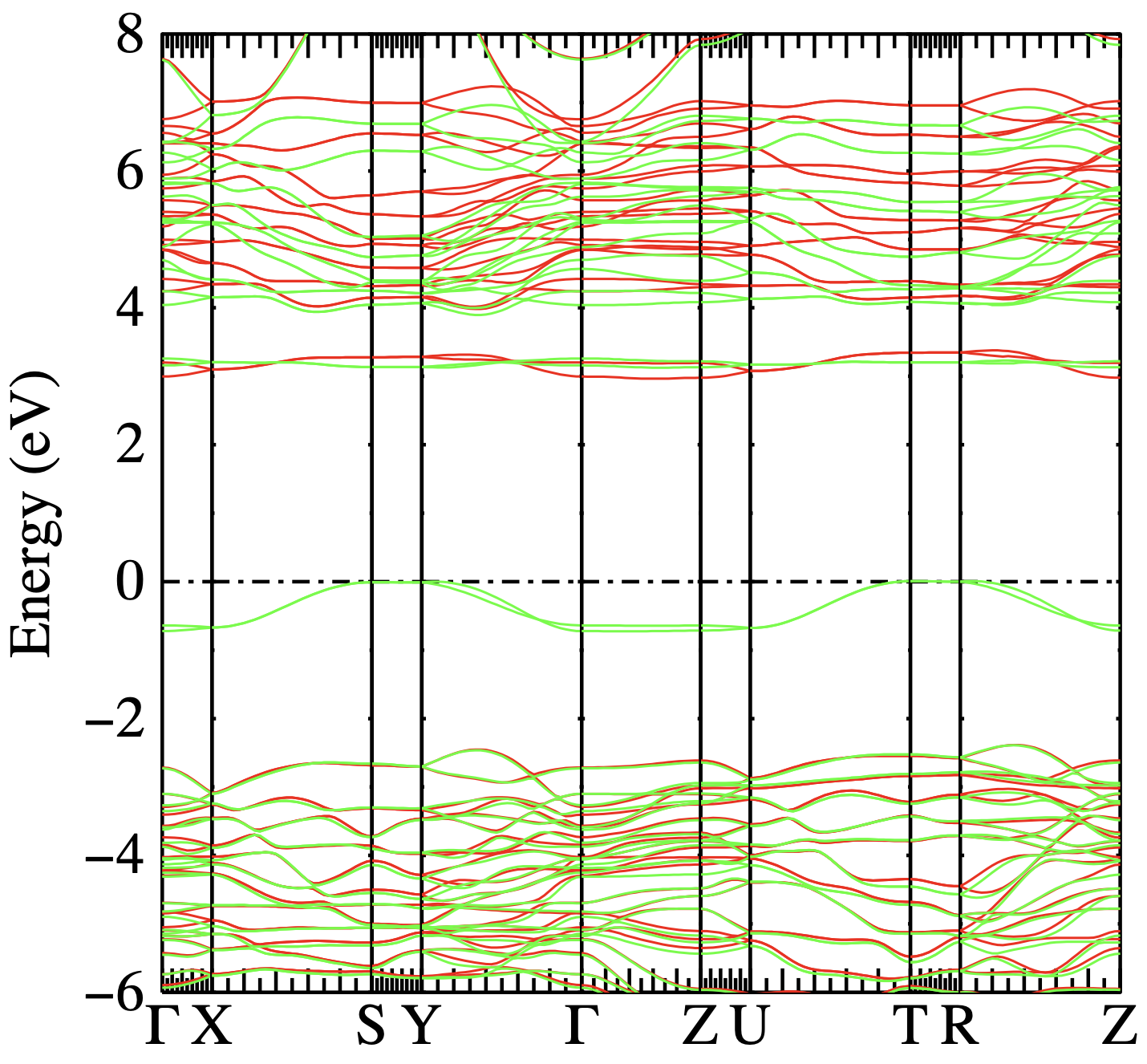}{}
        \caption{}
  \end{subfigure} 
        \begin{subfigure}{8cm}\centering
          \includegraphics[width=8cm]{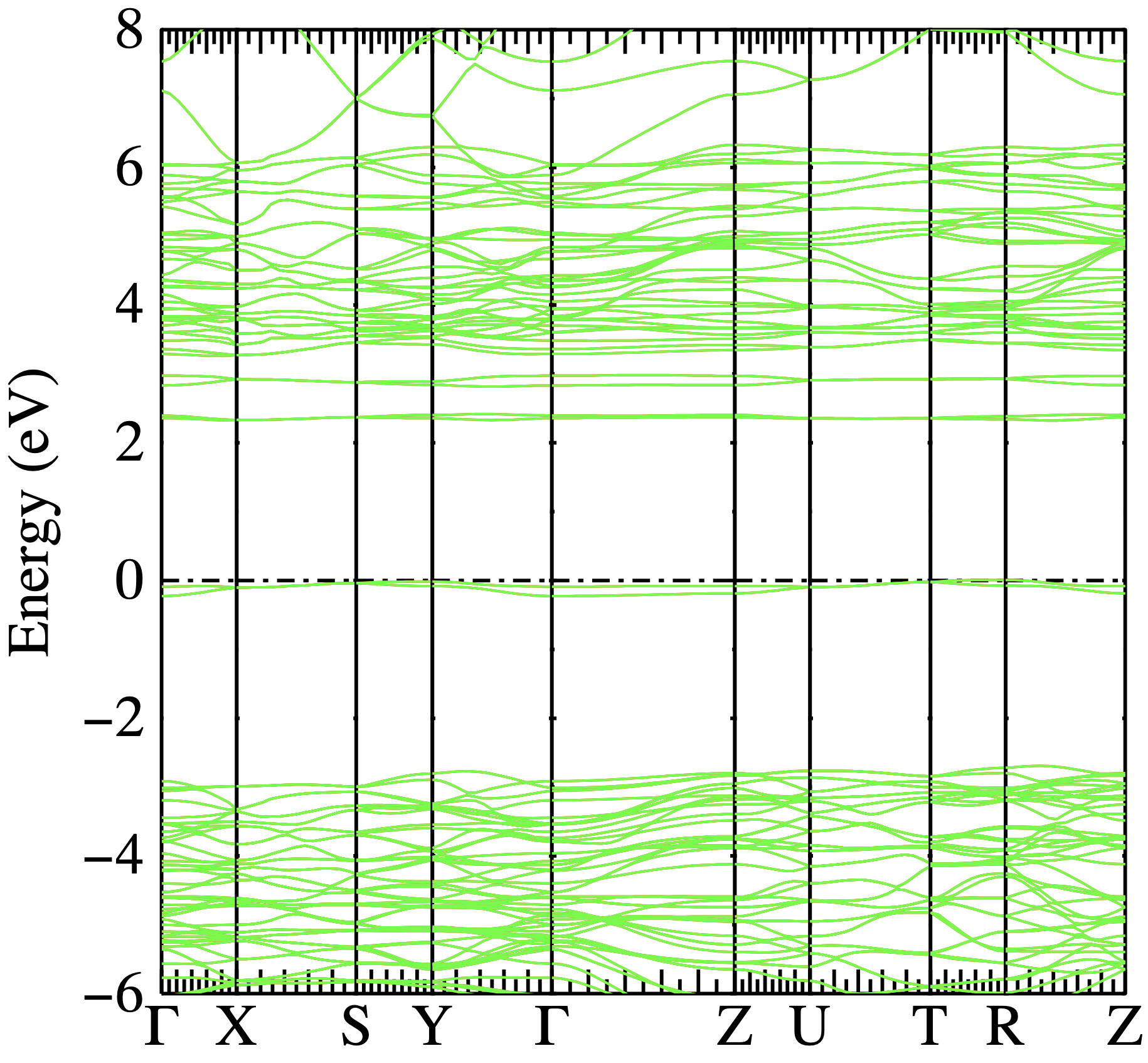}{}
             \caption{}
  \end{subfigure}
            \begin{subfigure}{8cm}\centering
              \includegraphics[width=8cm]{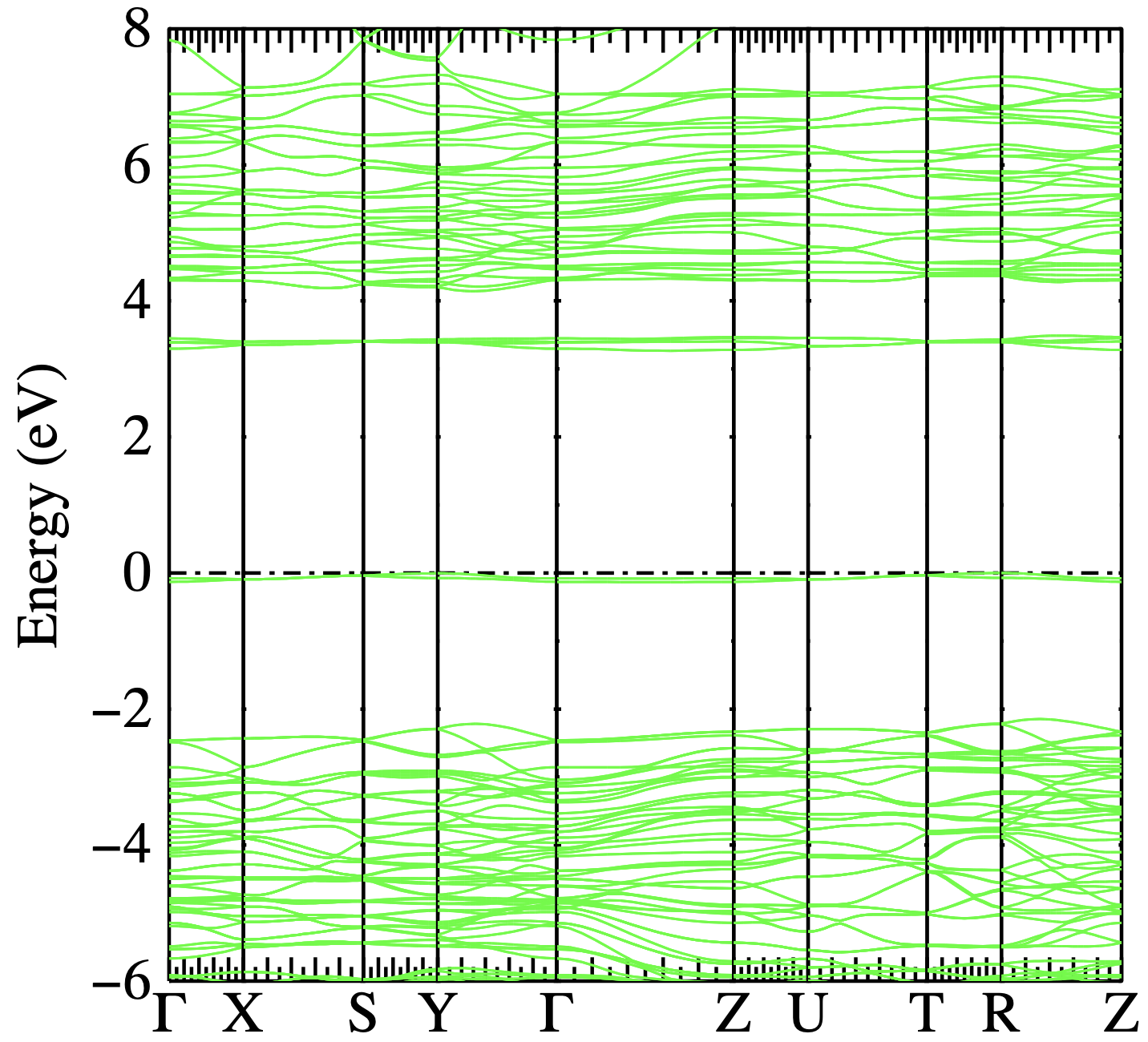}{}
                 \caption{}
  \end{subfigure}
   \caption{Energy band structures of LiV$_2$O$_5$ in (a) $Pmmn$ FM, (b),$P2_1mn$ FM, (c) $Pmmn$ AFM and (d) $P2_1mn$ AFM cases.\label{liv2o5bands}}
 \end{figure*}
  
  \subsection{Energy bands}
  We start by comparing the energy bands of the $Pmmn$ and the $P2_1mn$ structures  of LiV$_2$O$_5$ in Fig. \ref{liv2o5bands}. While at first sight looking very similar we may note that in the FM case, the lowest empty band has the same spin as the highest occupied split-off band  in the $Pmmn$ structure, while the band of opposite spin, in red, straddles the same spin band  in the $P2_1mn$ structure. In fact the splitting at $Y$ between the highest occupied majority spin band and the lowest empty minority spin band is 2.815  eV in $Pmmn$ but increases to 3.875  eV in the $P2_1mn$ case, while the lowest minority spin shifts from 3.090 eV to 3.770 eV.
  Thus the majority and minority spin bands are pushed closer together.
  Similarly in the AFM case the gap at $Y$ increases  from 2.434 eV to 3.396 eV.
  All of these bands are derived from the V-$d_{xy}$ orbitals, as analyzed in \cite{Bhandari15}, but the lower (occupied) ones have the same sign on the two V joined by a bridge O in the ${\bf a}$-direction, which we call ``rungs'' of the ladder compound,  while the higher (empty) ones of the same spin has opposite sign of the $xy$ orbitals.
When both are viewed as feeling an antibonding interaction second order interaction with the O-$p_y$ orbital of the bridge O, and therefore shifted up by $V_{dp\pi}^2/(\epsilon_d-\epsilon_p)$, 
the lower state forms a bonding combination in which this same interaction is cancelled out, while the upper is anti-bonding and thus is shifted by twice this amount. Alternatively, this cancellation can be viewed as resulting from the presence of a mirror plane perpendicular to the {\bf a}-axis passing 
through the bridge O$_b$.  The lower state is odd  with respect to the mirror and can therefore not interact with the $O_b$-$p_y$ while the upper state is even and can interact.
However, in the $P2_1mn$ case, the O$_b$ is displaced away from the
V$^{4+}$ and closer to the V$^{5+}$. Within this simple molecular model of the V-O-V bridge, it is easy to see why the gap increases when the mirror symmetry is broken. The change in potential on the two V essentially acts as an ionic contribution adding to the covalent interaction and thereby increases the gap. A detailed explanation is given in
the SI\cite{supinfo}.

  On the other hand, the next set of higher bands, which are of $xz$, $yz$ character are pushed up slightly higher so that the spin down band and the antibonding $xy$ derived band of majority spin become both detached from the main continuum of  higher $d$ bands. In the AFM configuration, the highest occupied and lowest few empty bands have notably smaller dispersion. This is because we only have non-zero hopping between orbitals of the same spin and in the AFM case, these amount to second nearest neighbors along the ${\bf b}$-direction.

  The band  structure of NaV$_2$O$_5$ in the FM and AFM $Pmmn$ and $P2_1mn$
  cases are given in the SI \cite{supinfo}. The $P2_1mn$ lowest conduction bands in that case show somewhat larger dispersion, the origin of which is discussed in SI.
 
 \subsection{Optical dielectric function}
 \begin{figure}
   \begin{subfigure}{8cm}\centering
     \includegraphics[width=8cm]{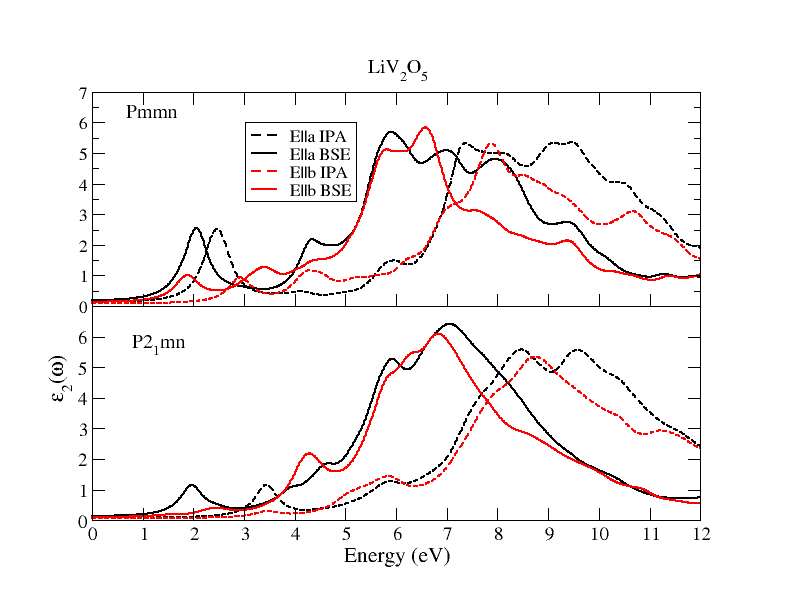}{}
     \caption{}
   \end{subfigure}
   \begin{subfigure}{8cm}
     \includegraphics[width=8cm]{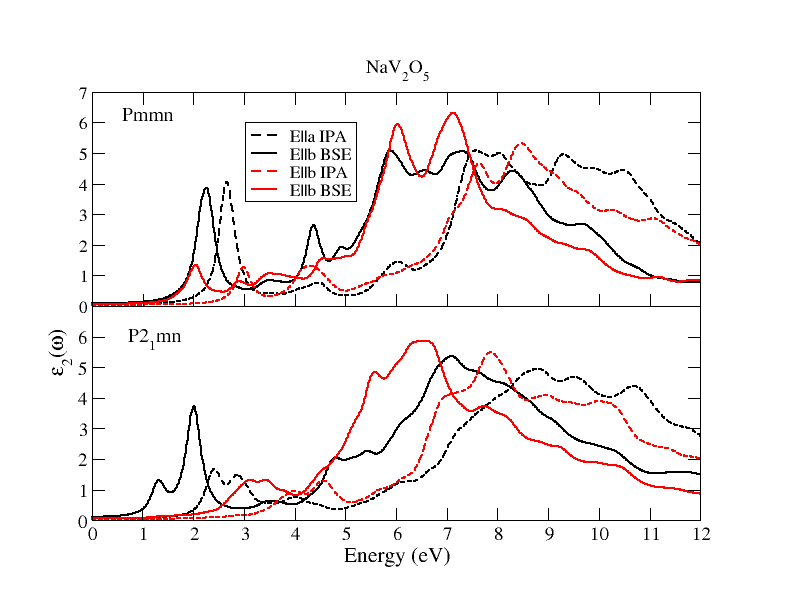}{}
     \caption{}
   \end{subfigure}
   \caption{Imaginary part of the dielectric function for $Pmmn$ and $P2_1mn$
     structures for (a)LiV$_2$O$_5$ and (b)NaV$_2$O$_5$.\label{figepspmmnp21mn}}
 \end{figure}
 Next, we compare the imaginary part of the dielectric functions $\varepsilon_2(\omega)$ between the two structures. Compared to pure V$_2$O$_5$, which
 is characterized by transitions from the occupied O-$2p$ band to the empty V-$3d$ bands, Li(Na)V$_2$O$_5$ shows additional features from $d-d$ transitions.
 These also show strong excitonic shifts as can be seen by comparing the independent particle approximation (IPA) with the BSE results. All systems show a marked anisotropy between the $a$ and $b$ polarizations.
  In the present work, we calculate these spectra using a rescaling\cite{Levine1989} of the optical matrix elements instead of explicitly attempting to evaluate the contributions of the $GW$ self-energy to the velocity matrix elements, which result from their non-locality.
There are notable differences between the two structures and between the Li and Na case. For example for $P2_1mn$ NaV$_2$O$_5$ we note a splitting of the lowest excitons related to the larger splitting of the first empty bands, seen in the band structure. We also note that the large excitonic effect here places the lowest optical transition closer to 1 eV. 
   
 \subsection{Magnetic properties}
\subsubsection{Magnetic moments}
  \begin{table}[h]
   \caption{Vanadium magnetic moments (in $\mu_B$), $m_t$ is the estimated true moment including an extrapolation beyond the sphere, $m_s$ is the moment within the muffin-tin sphere, and net valence charge $q_v$ (in $-e$).  \label{tabmom}}
   \begin{ruledtabular}
     \begin{tabular}{c|c|c|cc}
     structure  &&$Pmmn$&\multicolumn{2}{c}{$P2_1mn$}\\\hline
     site   & &V$^{+4.5}$ & V$^{4+}$&V$^{5+}$ \\\hline
     \multicolumn{5}{c}{LiV$_2$O$_5$}\\\hline
    FM &$m_t$&0.625        & 1.070 & 0.103 \\
    &$m_s$&0.471         & 0.784 & 0.066 \\
      & $q_v$& 1.833    & 1.712 & 1.553 \\ \hline 
      AFM &$m_t$&$\pm$0.608   & $\pm$1.061 & $\pm$0.087 \\
      & $m_s$&$\pm$0.460 & $\pm$0.777 & $\pm$0.060 \\
      & $q_v$&1.824      & 1.613      & 1.554 \\\hline
      \multicolumn{5}{c}{NaV$_2$O$_5$} \\ \hline
      FM & $m_t$ & 0.617& &  \\
      & $m_s$ & 0.443& &  \\
      & $q_v$ & 1.55 & & \\ \hline
      AFM &$m_t$& $\pm$0.586 &$\pm$0.889 & $\pm$0.254\\
      &$m_s$& $\pm$0.424 &$\pm$0.683& $\pm$0.174 \\
      &$q_v$& 1.55 & 1.99& 1.43 \\
     \end{tabular}
   \end{ruledtabular}
  \end{table}

  First we report the magnetic moments on V atoms and net valence electron charges per  muffin-tin sphere in Table \ref{tabmom}. While the net charge difference between the nominal V$^{5+}$ and V$^{4+}$ are quite small, the magnetic moments are strikingly different. This is because the spin density results from only the $d_{xy}$ derived
 bands as discussed in the previous section, while the net charge involves the contributions from all the V-$d$ participation in the bonding occupied states as well as the V-$4s$ and V-$4p$ contributions.  
 The  net magnetic moments however, clearly show the  charge disproportionation, which perhaps should rather be named a spin disproportionation. Nonetheless, the nominal  V$^{5+}$  still have  a small magnetic moment about one order of magnitude smaller than the V$^{4+}$ and in the $Pmmn$ structure the net magnetic moment is indeed close to 0.5 $\mu_B$. Small differences occur  between
 the absolute values of the moments in the FM and AFM configurations. We also note that a significant part of the V magnetic moment extends beyond its muffin-tin sphere. The radial spin density inside the muffin-tin sphere can be extrapolated with a smooth Hankel function beyond its sphere and allows one to estimate the true moment associated with a given atom. Turning to NaV$_2$O$_5$ we can see similar magnetic moments. However, we may notice that in the $P2_1mn$  the charge disproportionation is somewhat smaller. The ratio of the high to the small magnetic moment is only about 3.5 instead of an order or magnitude.

 \begin{table}
   
   \caption{Magnetic moments (in $\mu_B$) induced on O atoms in AFM structures.}
   \label{tabomom}
   \begin{ruledtabular}
     \begin{tabular}{ccc}
       site&$Pmmn$&$P2_1mn$\\ \hline
       \multicolumn{3}{c}{LiV$_2$O$_5$}\\ \hline
       O$_b$ & 0.067 & 0.026 \\
       O$_v$ & 0.048 & 0.067 (0.008)\footnote{Value in parentheses for $O_v$ connected to V$^{5+}$, other one to V$^{4+}$} \\ \hline
       \multicolumn{3}{c}{NaV$_2$O$_5$} \\ \hline
       O$_b$ & 0.065 & 0.048 \\
        O$_v$ & 0.042 & 0.064 (0.0005) \\
     \end{tabular}
   \end{ruledtabular}
\end{table}
 
 The magnetic moments induced on Li (0.00015 $\mu_B$) and the chain oxygen O$_c$ (0.00038 $\mu_B$) are negligible but the magnetic moment on the vanadyl and bridge oxygens are not.
In LiV$_2$O$_5$ $Pmmn$ the O$_v=0.048\mu_B$ inside its sphere and extrapolated beyond is 0.065 $\mu_B$ and is opposite to the V to which it is connected.
 The magnetic moment induced on the bridge O$_b=0.067\mu_B$ and is also opposite in sign to the moments on the V it links and well confined to its own sphere.
 While these magnetic moments are an order of magnitude smaller than on the V, we will see that they lead to strong V-O exchange interactions.
 Similarly, non-negligible magnetic moments occur on O$_v$ connected to the V$^{4+}$ and on O$_b$ in the $P2_1mn$ structure but the moment induced on the $O_v$ connected to a V$^{5+}$ is negligble. These induced moments for the different cases are summarized in Table \ref{tabomom}.

 \subsubsection{Exchange interactions}

 \begin{figure}
   \begin{subfigure}{8cm}\centering
     \includegraphics[width=8cm]{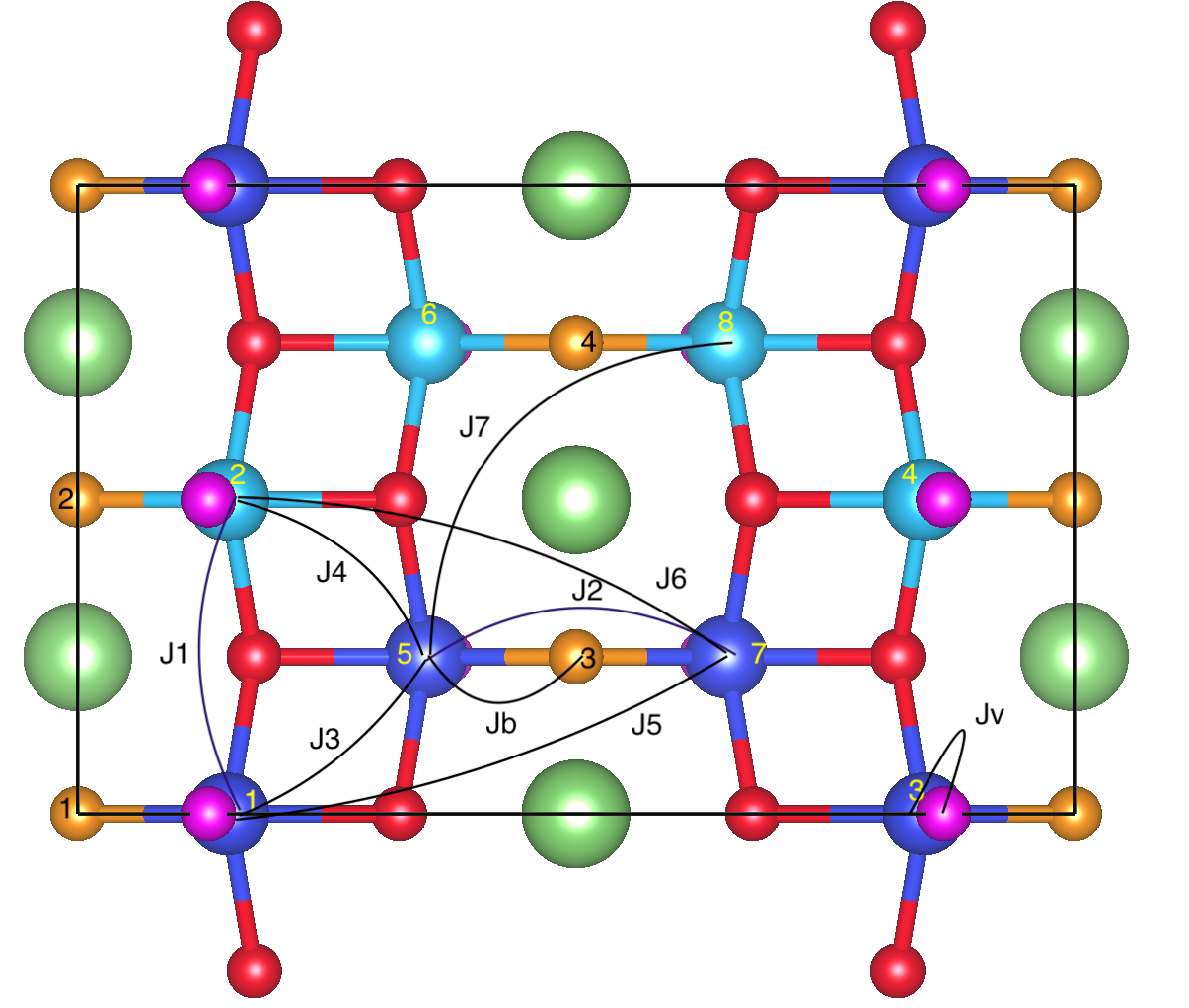}{}
     \caption{}
   \end{subfigure}
   \begin{subfigure}{8cm}
     \includegraphics[width=8cm]{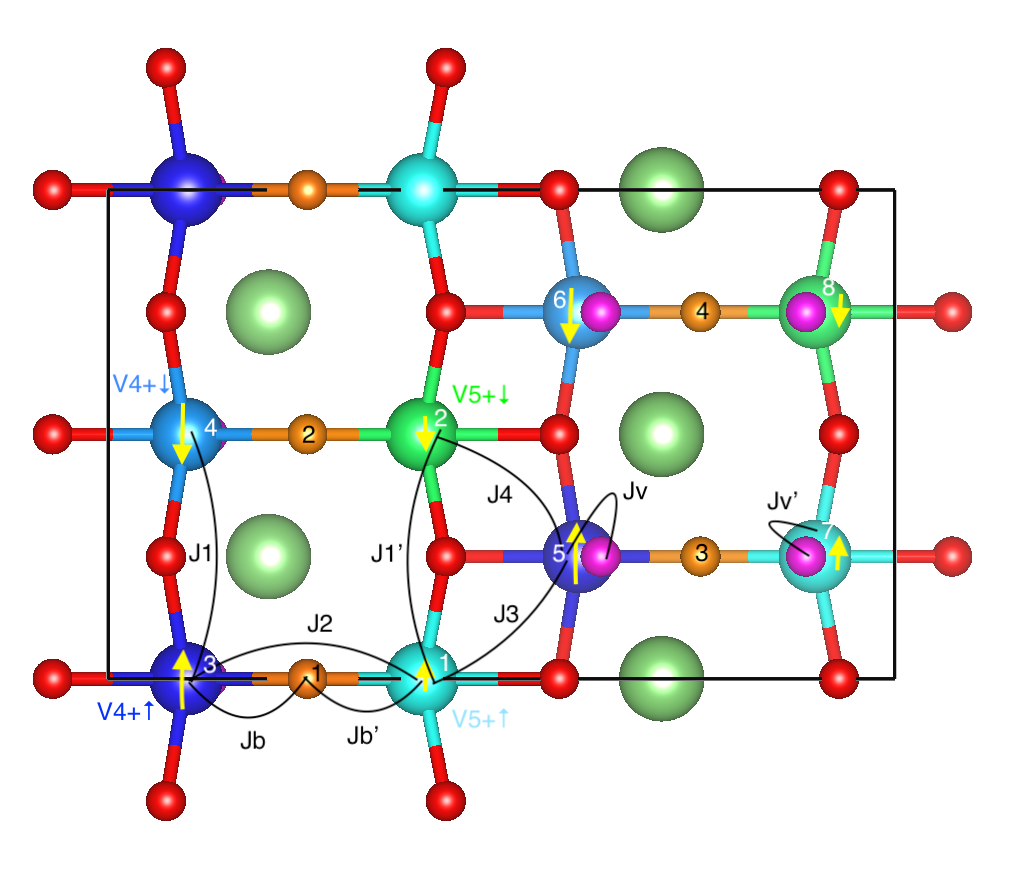}{}
     \caption{}
   \end{subfigure}
   \caption{Labeling of atoms and most important exchange interactions in (a) $Pmmn$ and (b) $P2_1mn$. Large green spheres, Li,  orange, bridge O, red chain O, magenta vanadyl O. In (a) V$\uparrow$ and V$\downarrow$ are dark blue and turqoise respectively, while in (b) 
  V$^{4+}\uparrow(\downarrow)$ are slightly different hues of dark blue and $V^{4+}\uparrow(\downarrow)$ are light blueish and greenish turqoise with large and small spins explicitly indicated as yellow arrows.}  This figures was made with the help of VESTA3 \cite{Vesta}.\label{figexchange}
 \end{figure}

 We label the exchange interactions as $J_{ab}^{\bf T}$ where the V-atoms in the cell are labeled $a,b$ and ${\bf T}=(n_1,n_2,n_3)$ labels the lattice vector difference between the unit cells of the $a$ and $b$ atom in reduced coordinates.
 Note that these are the $J_{ab}^c$ defined with respect to a classical Heisenberg Hamiltonian written in terms of unit vectors, as defined in Eq.(\ref{jclas}).
 In other words, they have already absorbed in them the size of the magnetic moments they are connecting. Additionally, we use a shortcut for some of the most important exchange interactions as simply $J_n$.
The labeling of the atoms and the most important exchange interactions are given in Fig. \ref{figexchange}. Note that the unit cell for $P2_1mn$ is shifted with repect to that of $Pmmn$ but we labeled the atoms in a corresponding manner.

By symmetry, we have 
 $J_1=J_{12}^{(0,0,0)}=J_{34}^{(0,0,0)}=J_{56}^{(0,0,0)}=J_{78}^{(0,0,0)}$ and each V has actually two equivalent interactions $J_1$ because $J_{12}^{(0,0,0)}=J_{12}^{(0,-1,0)}$, etc. For  
 $P2_1mn$, we use the same atom labeling but now, V1, V2, V7, V8 are V$^{5+}$ and have a small magnetic moment, while V3, V4, V5, V6 are V$^{4+}$ (with large magnetic moment) and  we need to distinguish the $J_{56}^{(0,0,0)}=J_{34}^{(0,0,0)}=J_1$ from
$J_{12}^{(0,0,0)}=J_{78}^{(0,0,0)}=J_1^\prime$. The interactions between V and the $O_v$ directly above or below it are labeled $J_v$ and with the bridge O $J_b$, but in $P2_1mn$ we distinguish the $J_v$ for a V$^{4+}$ from $J_v^\prime$ for V$^{5+}$ and the interactions of
V$^{4+}$ and V$^{5+}$ with the bridge O are labeled $J_b$ and $J_b^\prime$.

\begin{table}
  \caption{Exchange interactions $J_{ab}^{\bf T}$ in LiV$_2$O$_5$ in meV  as defined with respect to the classical Heisenberg Hamiltonian with unit vectors.
    All atoms treated as magnetic. For labels of atomic sites refer to Fig.\ref{figexchange}} \label{tabexchange}
   \begin{ruledtabular}
     \begin{tabular}{ccccccc}
       a&b&{\bf T}&$J_n$&$N_e$&\multicolumn{1}{c}{$Pmmn$}&\multicolumn{1}{c}{$P2_1mn$} \\ \hline
       V5&V6&(0,0,0)&$J_1$&2& -2.95& -6.85 \\
       V1&V2&(0,0,0)&$J_1^\prime$&2& & -0.16 \\
       V5&V7&(0,0,0)&$J_2$&1& 5.88 & -0.39\\
       V5&V1&(0,0,0)&$J_3$&1& -1.25 & -0.60 \\
       V5&V2&(0,0,0)&$J_4$&1& -1.33 & -0.60 \\
       V5&V3&(0,0,0)&$J_5$&2& 0.21  & 0.03 \\
       V5&V4&(0,0,0)&$J_6$&2& 0.20  & 0.04 \\
       V5&V8&(0,0,0)&$J_7$&2& -0.82 & -0.06 \\
       V5&O$_v^5$&(0,0,0)&$J_v$&1& -14.47 & -31.81\\
       V5&O$_b^3$&(0,0,0)&$J_b$&1& -7.19 & -2.08 \\
       V1&O$_v^1$&(0,0,0)&$J_v^\prime$ &1&  & -0.57\\
       V7&O$_b^3$&(0,0,0)&$J_b^\prime$ &1&  & -0.98 \\
       V5&V5&(0,0,1)& &1& -0.41 & -0.98 \\
       V5&V5&(0,1,0)& &2& -0.07 & -0.15 \\
       V5&V7&(0,0,1)& &2& -0.30 & -0.39 \\              
     \end{tabular}
   \end{ruledtabular}
\end{table}

\begin{figure}
  \begin{subfigure}{8cm}\centering
    \includegraphics[width=8cm]{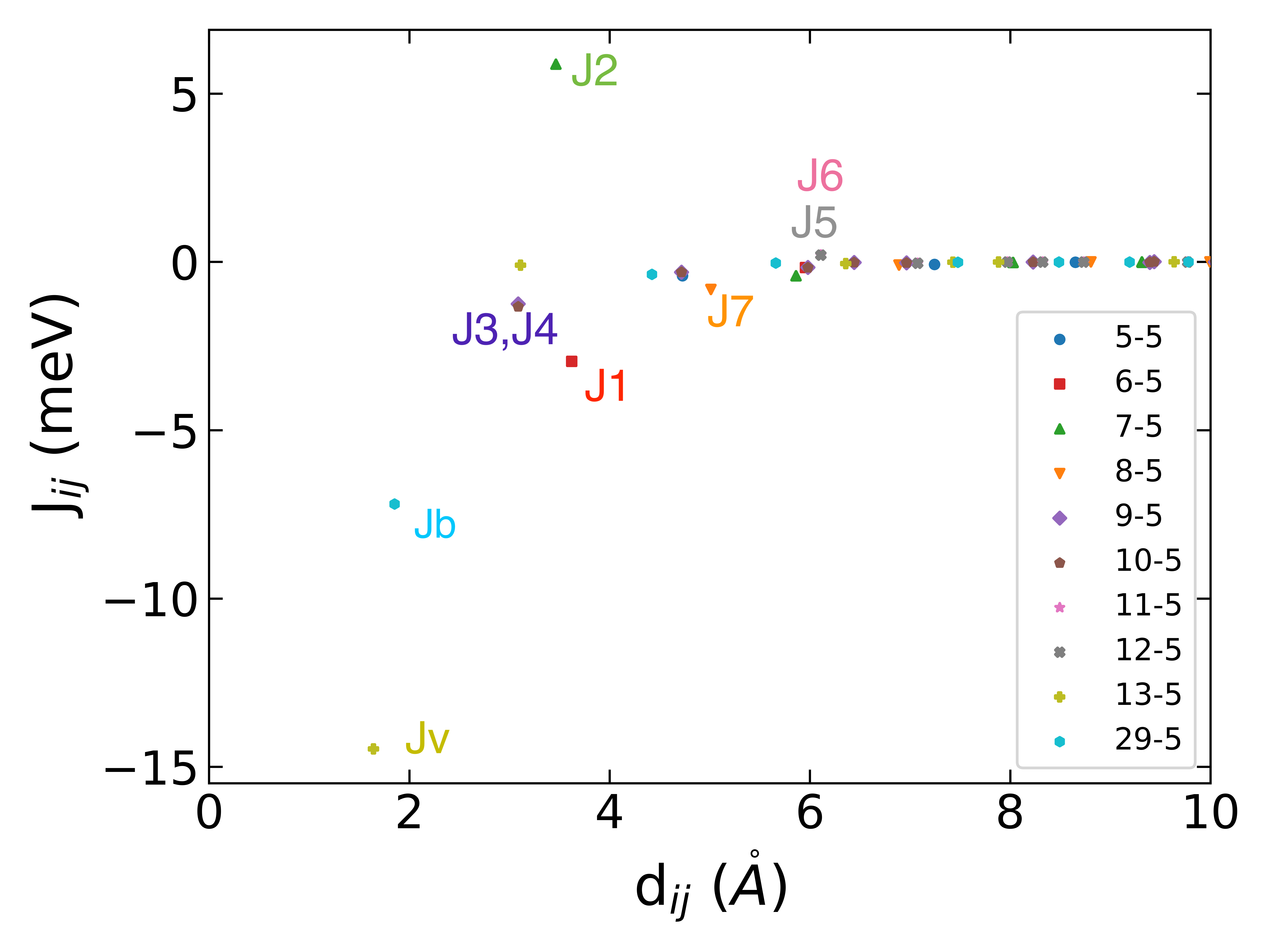}{}
    \caption{}
  \end{subfigure}
  \begin{subfigure}{8cm}\centering
    \includegraphics[width=8cm]{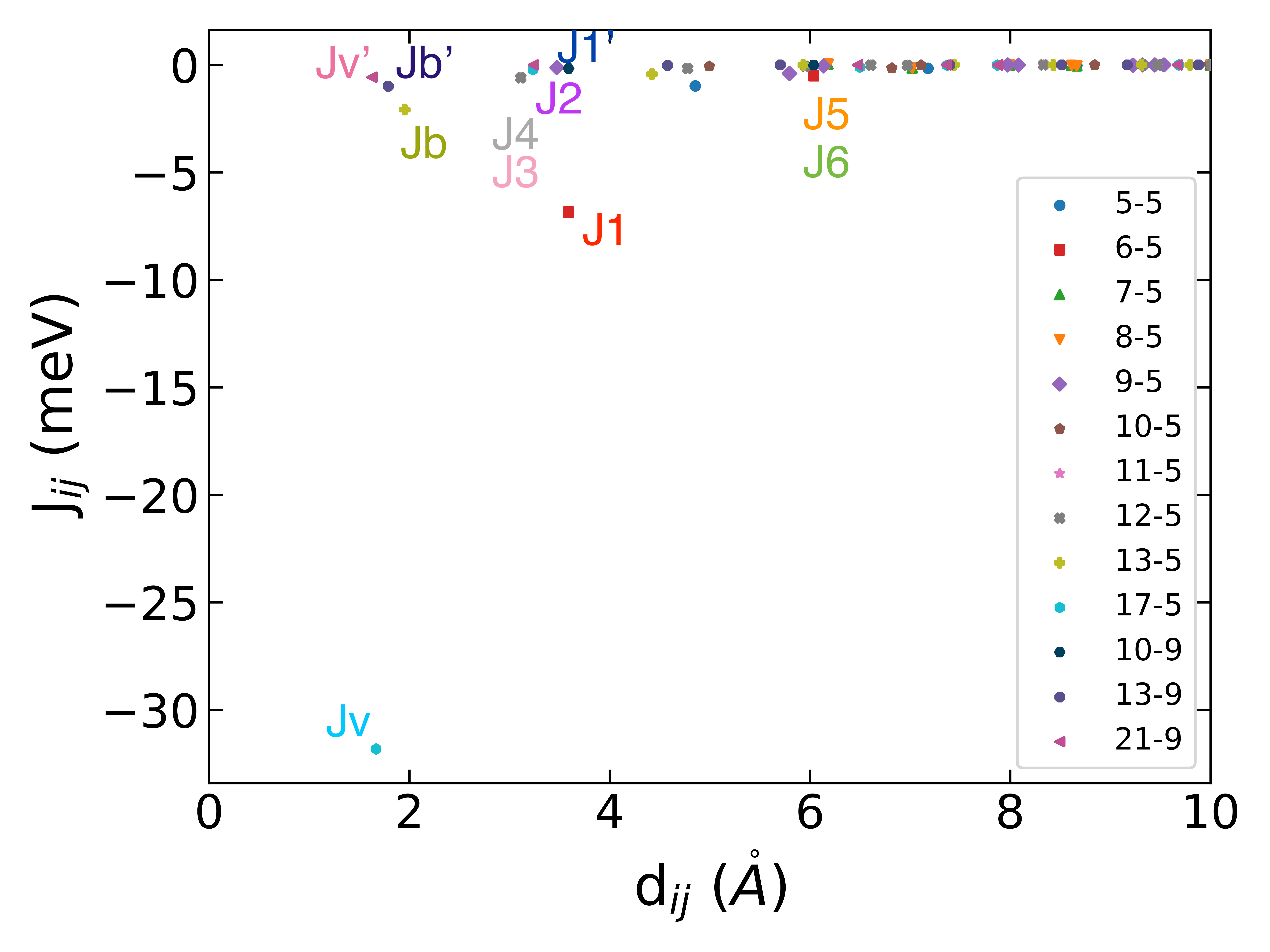}{}
    \caption{}
  \end{subfigure}
    \caption{Exchange interactions as function of distance for various pairs of atoms. (a) $Pmmn$  Note that here atoms 5-12 correspond to V1-V8, atom 13 is O$_v$ sitting on top of atom V1 and atom 29 is the O$_b$ between V1 and V3, referring to Fig. \ref{figexchange}; (b) $P2_1mn$ Here atom 5,6,9,13,17 correspond to V3,V4,V1,O$_b^1$,O$_v^3$ in   Fig. \ref{figexchange}.} \label{figj-d}
 \end{figure}
 
Table \ref{tabexchange}
lists the most important exchange interactions obtained from the AFM reference state, which has indeed the lowest total energy.  The number of nearest equivalent exchange interactions to a given one is denoted by $N_e$. In this calculation, all atoms are treated as magnetic, including the extremely low moment Li and O$_c$. 

Fig. \ref{figj-d} shows the exchange interactions as function of distance for
various pairs of atoms. One can see that beyond $\sim$6 \AA\ the $J_{ij}^{\bf T}$ for a given pair of atoms (i,j) the interactions become negligible and in fact
only $J_v$, $J_b$, $J_1$,$J_2$ are important. 

 We can see that the nearest neighbor interaction along the ${\bf b}$-chains is negative, indicating antiferromagnetic coupling and is more than twice as large in the $P2_1mn$ structure for the high moment atoms and an order of magnitude smaller for the low moment atoms. This makes sense because these interactions incorporate the factor $|S|^2$. The exchange interaction $J_2$ is strongly ferromagnetic as expected in $Pmmn$ since it actually  corresponds to  a single electron spread over two V. The $J_2$ in $P2_1mn$ on the other hand is strongly reduced because this is now an interaction between a large and small moment and is even seen to switch sign.   
 The interactions $J_3$ and $J_4$ are smaller and both negative and close to each other.
 They are lower for $P2_1mn$ because then they are between large and small moment atoms. 
 We may note that $J_4$ being negative agrees with the reference system spins being opposite but $J_3$ being negative is opposite to the spins being parallel. This indicates geometric frustration in the spin system. The system cannot satisfy both of these interactions to being  antiferromagnetic at the same time.
 The interaction $J_7$ between atoms across the large interstitial site where Li is located is also antiferromagnetic and still significant although smaller than the main interactions by one order of magnitude.  The rather long-range interactions such as $J_5$ and $J_6$ are already fairly small but are indicated here because there are equivalent interactions with a neighboring cell along the ${\bf a}$-direction which will lead to dispersion of the spin waves along this direction. 
 The table  is not exhaustive and other interactions of similar magnitude of a few 0.1 meV exist. For our subsequent analysis of the critical temperatures, we focus on  the strongest interactions but our numerical calculations include all of them.  We may also note that interactions between adjacent layers tend to be negative.
 An example is the interaction between V5 and V5 in adjacent layers but also between V7 and V5.
 The fact that the interactions between layers is negative indicates that the reference state
 considered here which has the same spin directions in each layer may not be the ground state and opposite spin from layer to layer would lower the total energy. This would then require a 64 atom cell and has yet to be further studied. The actual ground state of the system has in fact been found to have charge disproportionation involving an ordering with multiple layers and the ordering of the spins in this system has not yet been definitively identified.

 Quite surprisingly, the strongest exchange interactions are actually between the vanadyl O$_v$ and its adjacent  vanadium this interaction is antiferromagnetic, which is consistent with our finding that the moments on V and its adjacent O$_v$ are antiparallel. It is found here to be $-14.47$ meV in $Pmmn$ and more than double this value in $P2_1mn$ for the V$^{4+}$. The exchange interaction with the bridge O$_b$ is also large and negative. This is especially surprising because these interactions between unit vectors in fact include the magnetic moment factors
 and the magnetic moments on these oxygens is an order of magnitude smaller than on V.
 This is a further indication of the distributed nature of the spin in this overall $S=1/2$ system.
 Not only is the spin distributed equally over two V atoms in $Pmmn$,  it also spreads  to the
 corresponding vanadyl O$_v$ and is also significant on the O$_b$ in between the  vanadiums. The exchange interactions in NaV$_2$O$_5$ are similar to those in  LiV$_2$O$_5$ but the charge disproportionation is less pronounced and hence the interactions involving the nominal V$^{5+}$ or small magnetic moment V are somewhat larger. They are given in SI and were discussed in more detail in \cite{Lambrechtnav2o5}

 \subsubsection{Critical temperatures}
 Next, we discuss the critical temperature in the mean field approximation.
 First, numerically diagonalizing the ${\bf J}_0$ matrix of Eq.  \ref{eq:tc}
 including all atoms in the cell, 
 we find for $Pmmn$ of LiV$_2$O$_5$ a $T_N=187.3$ K and for $P2_1mn$ 293 K.
The results also depend on how many atoms we treat as magnetic atoms. If we 
treat V and the O$_v$ and O$_b$ as magnetic atoms we find a mean field
temperature of 173 K.
 Both of these are significantly smaller than the observed $T_{max}\approx 320\pm50$ K
for NaV$_2$O$_5$.  No experimental data are available for $\alpha$-LiV$_2$O$_5$ N\'eel temperature but because the exchange interactions are similar, we may assume their critical temperatures also are close to each other. 
However, if we recall that we describe in fact a nominal $S=1/2$ system
 even though we here calculate it with distributed moments over several sites,
 and that for small spins the quantum mechanical correction factor $(S+1)/S$, here a factor 3 is important, we estimate $T_N^{QMFA}=562$ K for $Pmmn$ and 879 K for $P2_1mn$.
 As is well known, mean field theory tends to overestimate critical temperatures.
 At this point, it is not clear by how much but for low dimensional systems like a square lattice the actual $T_c\approx0.56T_c^{MF}$. Applying this factor we obtain 315 K for $Pmmn$ which is consistent with the data, while the value for $P2_1mn$ of 492 K appears to overestimate $T_N$.
 Because the $P2_1mn$ has lower total energy 
one expects locally a $P2_1mn$ like distortion with charge disproportionation, but with significant disordered fluctuations which on average 
give a $Pmmn$ structure in XRD. Nonetheless, the fluctuations in some sense spread out the moment more over the two V across the bridge and this seems to be an important factor in reducing the $T_N$ compared to the more ordered $P2_1mn $ structure.

To gain further insight in the role of the different exchange interactions,  we now consider different models keeping only a limited set of $J_{ij}$. 
At first, one might have assumed that we only need to keep interactions between the large V magnetic moments. 
 Including only the most important interactions $J_1-J_4$, the matrix of Eq.(\ref{eq:tc}) for $Pmmn$ becomes 
\begin{equation}
  \bm{J}^0=\left(\begin{array}{cccccccc}
   0&J_2&2J_1&0&0&J_3&0&J_4\\ 
   J_2&0&0&2J_1&J_3&0&J_4&0\\
   2J_1&0&0&J_2&0&J_4&0&J_3\\
   0&2J_1&J_2&0&J_4&0&J_3&0\\
   0&J_3&0&J_4&0&J_2&2J_1&0\\
   J_3&0&J_4&0&J_2&0&0&2J_1\\
   0&J_4&0& J_3&2J_1&0&0&J_2\\
   J_4&0&J_3&0&0&2J_1&J_2&0
  \end{array}\right).
\end{equation}
Because of the symmetry, each row in this matrix is a permutation of the first row and the matrix 
can be diagonalized analytically (using Mathematica) and has as highest eigenvalue
$-2J_1+J_2-J_3+J_4$  which gives a $T_c^{MF}=90$ K. This is almost a factor two smaller than the calculation including O$_v$ and O$_b$ indicating the importance of the latter. 
We may note that $J_3$ an $J_4$ are close to each other in value and almost cancel each other out.  This indicates the frustration between these interactions. Neglecting them, $T_N^{MF}\approx (2/3k_B)(J_2-2J_1)$, when only keeping the V magnetic moments. Note that compared to the basic 1D chain model with $T_N^{MF}=(2/3k_B) 2|J_1|$, the presence of   
$J_2$ provides   a major contribution to $T_N$. However, it is still significantly lower than the full calculation.

For the $P2_1mn$ case the matrix becomes
\begin{equation}
  \bm{J}^0=\left(\begin{array}{cccccccc}
   0&J_2&2J_1&0&0&J_3&0&J_4\\ 
   J_2&0&0&2J_1^\prime&J_3&0&J_4&0\\
   2J_1&0&0&J_2&0&J_4&0&J_3\\
   0&2J_1^\prime&J_2&0&J_4&0&J_3&0\\
   0&J_3&0&J_4&0&J_2&2J_1&0\\
   J_3&0&J_4&0&J_2&0&0&2J_1^\prime\\
   0&J_4&0& J_3&2J_1^\prime&0&0&J_2\\
   J_4&0&J_3&0&0&2J_1^\prime&J_2&0
  \end{array}\right)
\end{equation}
The highest eigenvalue is then
$-J_1-J_1^\prime+\sqrt{(J_1-J_1^\prime)^2+J_2^2+J_3^2+J_4^2+2J_2J_3-2J_3J_4-2J_2J_4}$. Since $J_1^\prime$ and $J_2$ are negligible, and even $J_3$ and $J_4$ are small, we can approximate it as
$|J_1|+\sqrt{J_1^2+J_3^2+J_4^2}\approx 2|J_1|$. However, the $J_1$ is now larger by a factor 2.3 and we obtain $T_N\approx 105 $ K. In the end, this is close to what we had for $Pmmn$ if we keep only V interactions. Essentially, we recover here the isolated chain limit. So, by charge disproportionating the spin and isolating it on one chain we increased the interactions along the chain but we lost the ferromagnetic coupling $J_2$ keeping the spins on both V parallel and only a modest change results in the critical temperature.

Now, if we include the $O_v$ and $O_b$ exchange interactions, we need to
diagonalize a $20\times20$ matrix. It becomes a bit cumbersome to write this matrix out here but we find it can still be diagonalized by Mathematica and gives as highest eigenvalue
\begin {equation}
  j_{max}=\frac{1}{2}\left[-2J_1+J_2+\sqrt{(-2J_1+J_2)^2+8J_b^2+4J_v^2}\right].
\end {equation}
  This gives a $T_N^{MF}\approx189$ K, close to the value obtained from
  diagonalizing the full matrix including all interactions.
  The exchange interactions between O and V are thus found to be crucial in establishing the correct scale of the  critical temperature.
  For $P2_1mn$ the same equation applies if we just neglect $J_2$ and we then obtain 277 K, close to the numerical value obtained using all interactions, which is 293 K.
  So, the approximate analytical treatment including only the main interactions is well
  justified. 
  
Using the RPA for these systems is somewhat tricky. A large mesh of $10\times10\times10$ is used and various values of the analytic continuation parameter $\epsilon$ are used and only small mixing is used in the iterations to ensure the final fixed point is not deviating too much from the initial mean field solution. Also the $T_c$ obtained from each of the sublattices shows still deviations from the V and O sublattices of order a few K. With these cautions, we obtain a $T_N^{RPA}$ of about 140$\pm 5$ K when including O$_v$ and O$_b$ as magnetic atoms in addition to V. Multiplying by the factor 3 this gives
420 K, for the $Pmmn$ structure.
This is larger than for NaV$_2$O$_5$ where we obtained 360 K in this manner, which is in reasonable agreement with the experimental value of 320$\pm50$ as reported by Carpy \etal \cite{Carpy72}.
Consistent with the higher exchange interactions in the Li than the Na case, this predicts a somewhat higher critical temperature for LiV$_2$O$_5$.

   Interestingly, the oxyfluorides NaV$_2$(O$_{1-x}$F$_x$)$_5$ even for very small $x$ were found to have a much smaller $T_N\approx 140-150$ K and were found to have the $Pmmn$ structure. Our results indicate that this difference  is not due to the difference in the symmetry of the structures. We find only a relatively small increase in the MFA $T_N$ from  $Pmmn$ to $P2_1mn$ and it is mainly due to the localization of the magnetic moment on one V instead of two but which also increases its exchangee interaction with the vanadyl O attached to it, both of which increase
   by more than a factor two. We speculate that substitution of O by F would dope the system with electrons  in the lowest conduction band. This could lead to additional carrier mediated ferromagnetic interactions, which would counteract the antiferromagnentism and thus possibly reduce the antiferromagnetic N\'eel temperature.  This tentative explanation requires further study beyond the scope of the present paper.

   \subsubsection{Spin waves}
\begin{figure}[ht]
     \includegraphics[width=8cm]{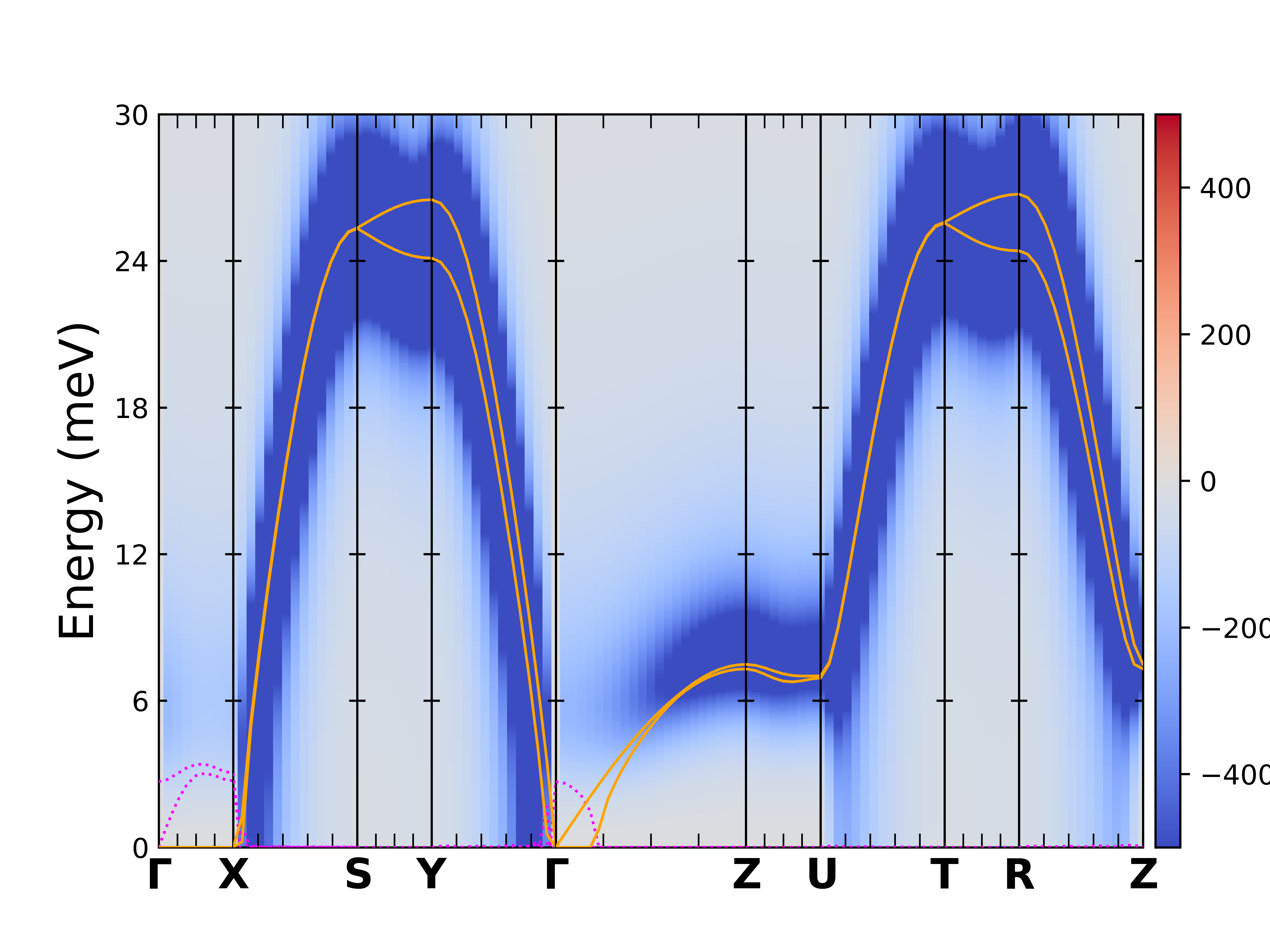}
     \caption{Spin wave dispersion (yellow solid lines of dotted pink if imaginary) along with ${\rm Im}[{\rm Tr}[\chi^{+-}({\bf q},\omega)]]$ as heatmap
       for $P2_1mn$ LiV$_2$O$_5$ including only the four V$^{4+}$ as magnetic atoms.} \label{fig:swp21mn-v4}
\end{figure}
    If we include all the 32 atoms in the unit cell, we obtain 16 positive and 16 negative frequency spin waves of equal magnitude and oppostie sign. To gain insight we start by considering the $P2_1mn$ case and including only the four V$^{4+}$ atoms with large magnetic moments $M_{{\rm V}^{4+}}=0.777$ $\mu_B$. The resulting spin wave eigenvalues are shown in Fig.\ref{fig:swp21mn-v4}. There is only one main interaction in this case, $J_1=-2.44$ meV  between nearest neighbors along the chain.  In that case the spin wave along $\Gamma Y$ for a single chain 
   is given by $\omega_{\bf q}=\pm|2MJ_1\sin{(q_yb/2)}|$. Note that $J_1$ here is referred to the Heisenberg Hamiltonian written in terms of spins  $S_i=M_i/2$ while the values in Table \ref{tabexchange} are with respect to unit vectors and incorporate the magnitude of $S$. Thus the values in the table need to be divided by $S^2$ with $S=M/2$ from Table \ref{tabmom}, giving   $J_1/|S|^2=-16$ meV. At the zone boundary we then obtain 25 meV. This is seen to agree well with Fig. \ref{fig:swp21mn-v4} if we keep in mind that there are two weakly interacting chains per unit cell, which slightly splits this mode in two modes. The dispersion along other directions results from weaker long range interactions. We note that the eigenvalues along $\Gamma X$ and one of them along $\Gamma Z$ are imaginary, which indicates the reference structure is unstable with respect to these spin waves and this results from the negative exchange interactions between atoms in the corresponding directions. Fig. \ref{fig:swp21mn-v4} also shows
    ${\rm Im}[{\rm Tr}[\chi^{+-}({\bf q},\omega)]]$ and this peaks at the poles of the $\chi^{+-}$ and tracks the spin wave eigenvalues.

   Continuing to add complexity we next consider the $Pmmn$ structure
   keeping only the V atoms.
   We then need to consider the two chains forming a ladder with rungs
   centered on the bridge O$_b$.  Again, let us focus on a single  ladder first.
   The matrix giving the spin wave dispersions is then give by 
\begin{widetext}
\begin{equation}
  \bar{\bf J}=\left(\begin{array}{cccc}-J_2+2J_1& J_2& 2J_1\cos{(q_yb/2)}& 0\\
    J_2& -J_2+2J_1&0&  2J_1\cos{(q_yb/2)}\\
    2J_1\cos{(q_yb/2)}&0& -J_2+2J_1& J_2\\
    0&  2J_1\cos{(q_yb/2)}& J_2& -J_2+2J_1
  \end{array}\right).
\end{equation}
\end{widetext}
The generalized eigenvalue probem $\det[\bar{\bf J}-\omega{\bf M}^{-1}]=0$
then gives the spin wave dispersions 
\begin {eqnarray}
  \omega_1({\bf q})&=&\pm|2MJ_1\sin{(q_yb/2)}|,\nonumber \\
  \omega_2({\bf q})&=&\pm 2|M|\sqrt{J_1^2\sin^2{(q_yb/2)}+J_2(J_2-2J_1)}\label{eq:sw1}
\end {eqnarray}

\begin{figure}[h]
  \includegraphics[width=8cm]{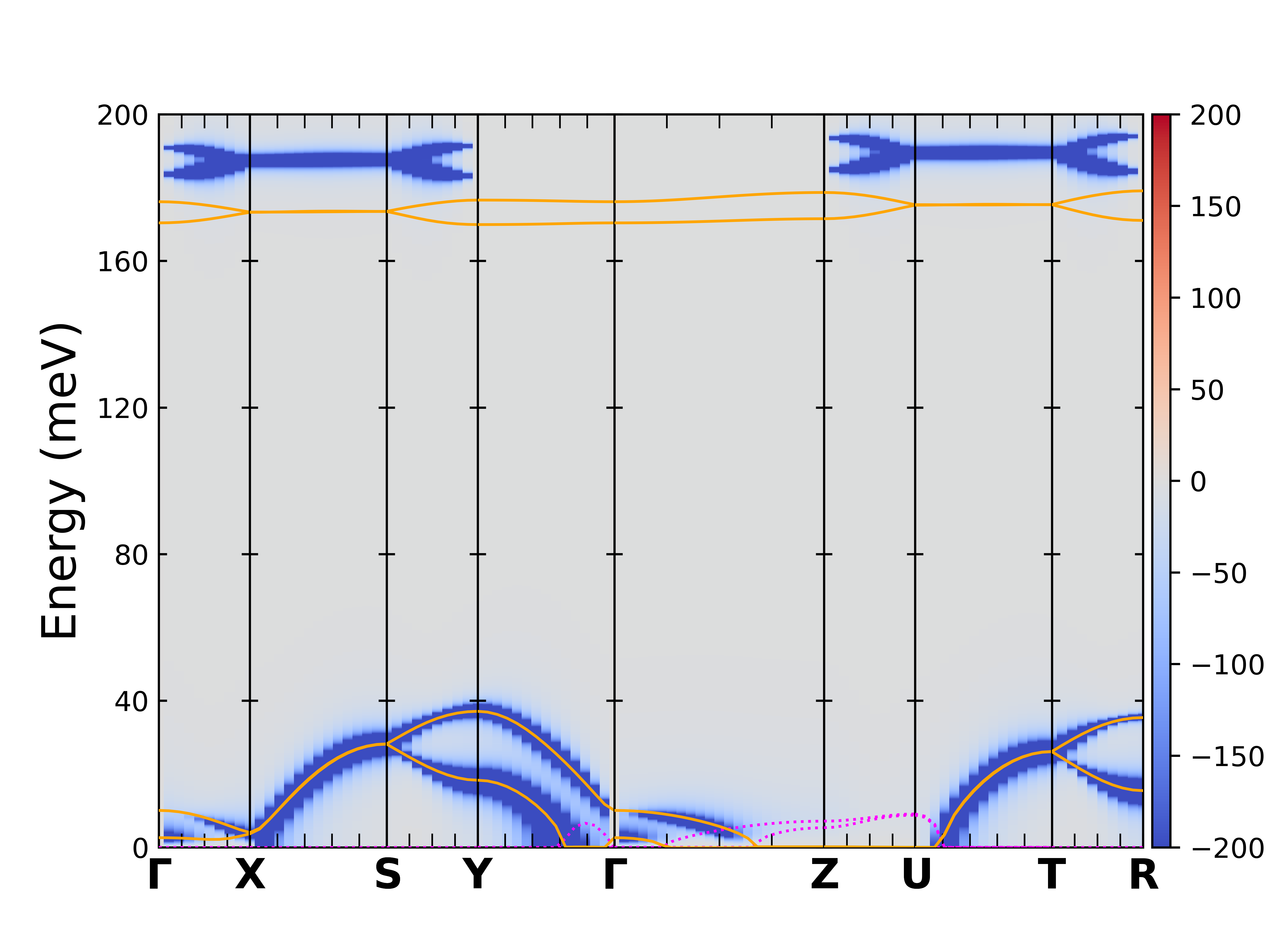}
  \caption{Spin wave dispersion and  ${\rm Im}[{\rm Tr}[\chi^{+-}({\bf q},\omega)]]$ of $Pmmn$ LiV$_2$O$_5$ using only V as magnetic atoms.}
  \label{fig:sw-liv2o5-pmmn-v}
\end{figure}
The eigenvectors for the first eigenvalue correspond to the spins on atoms 1 and 2 across a rung precessing in the same direction, while the second eigenvalue corresponds to the spin precessing in opposite direction. It is thus an optic spin wave instead of an acoustic one. There still only is dispersion along the
chain direction in this model. Using $J_1=-1.05$ eV and $J_2=8.26$ eV obtained in this approximation, the equation predicts an optic spin wave about 180 meV and maximum of the acoustic spin wave at $\sim20$ meV. We compare this model with the corresponding numerical calculation in Fig. \ref{fig:sw-liv2o5-pmmn-v}, which indeed shows two spin waves near 170 meV and a maximum of the lowest spin wave at 18 meV.
The model including only V atoms can be further extended to include all 8 atoms and including exhange interactions $J_3-J_6$. We do this in the appendices. First in Appendix \ref{appa} we show that the problem can be reduced to one spin sublattice and then we apply it to the system of 8 vanadiums per cell providing analytic equations for the eigenvalues in Appendix \ref{appb}.

\begin{figure*}
  \begin{subfigure}{8cm}\centering
    \includegraphics[width=8cm]{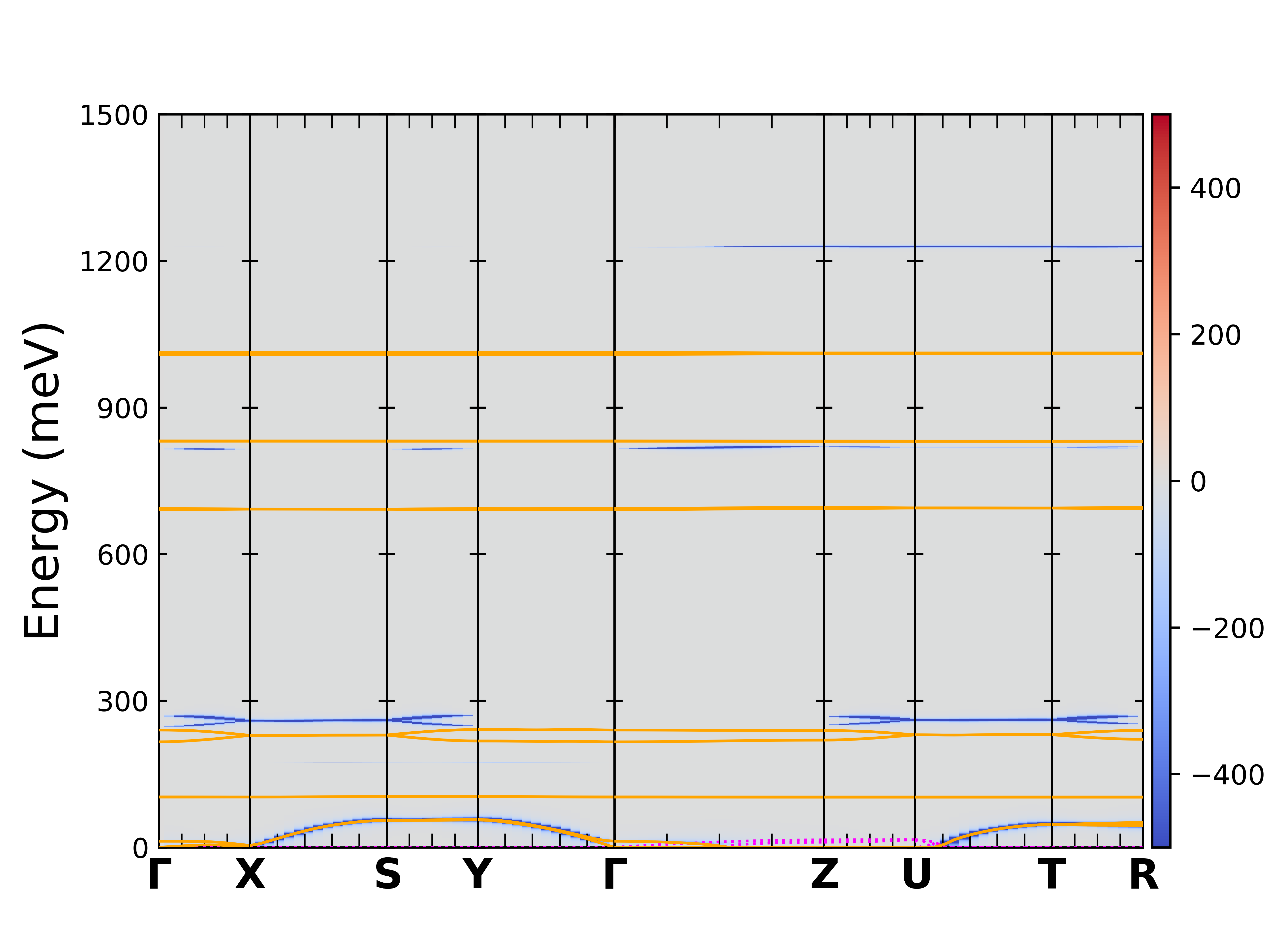}{}
    \caption{}
  \end{subfigure}
 \begin{subfigure}{8cm}\centering
   \includegraphics[width=8cm]{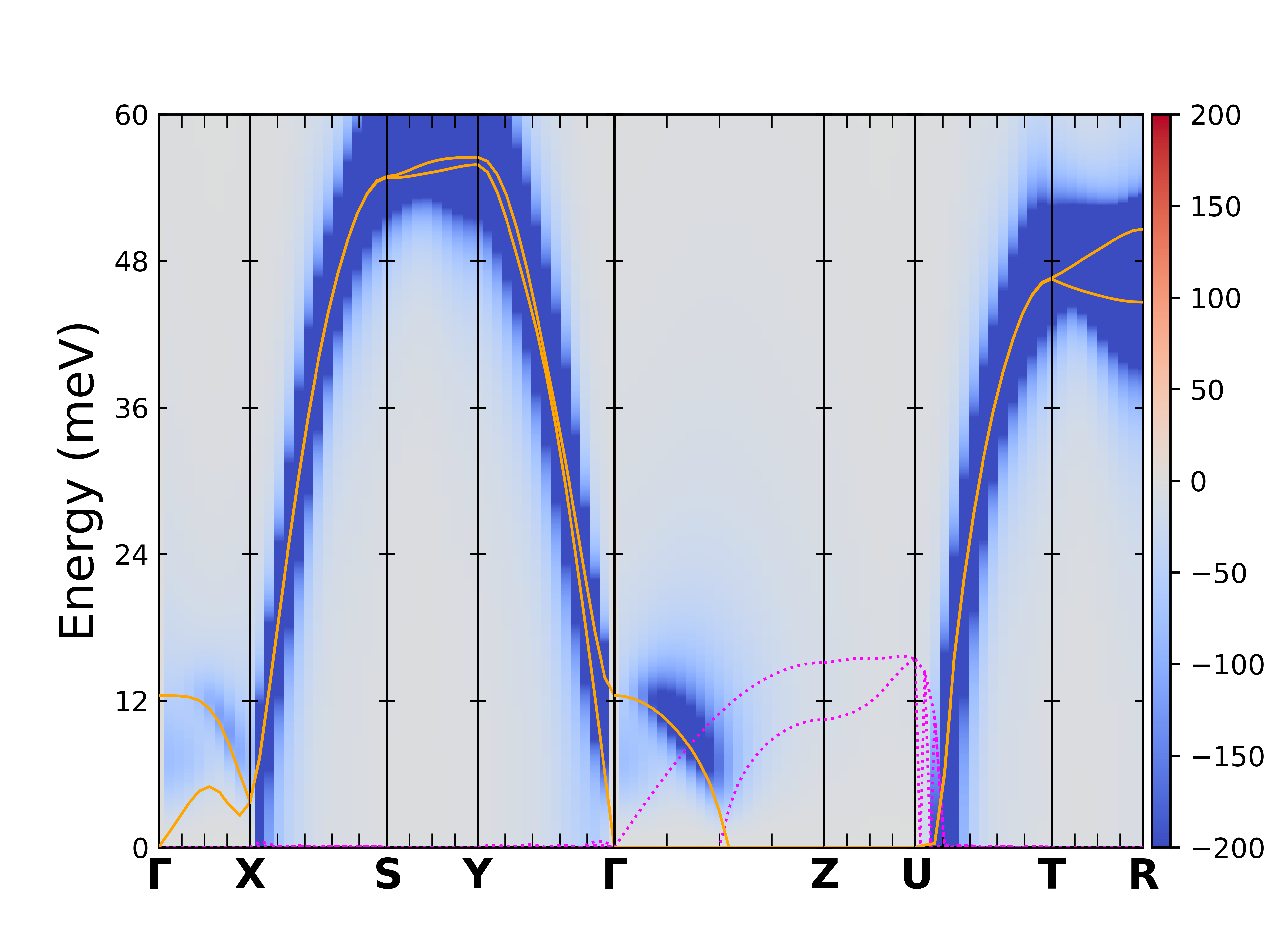}{}
   \caption{}
 \end{subfigure}
  \begin{subfigure}{8cm}\centering
    \includegraphics[width=8cm]{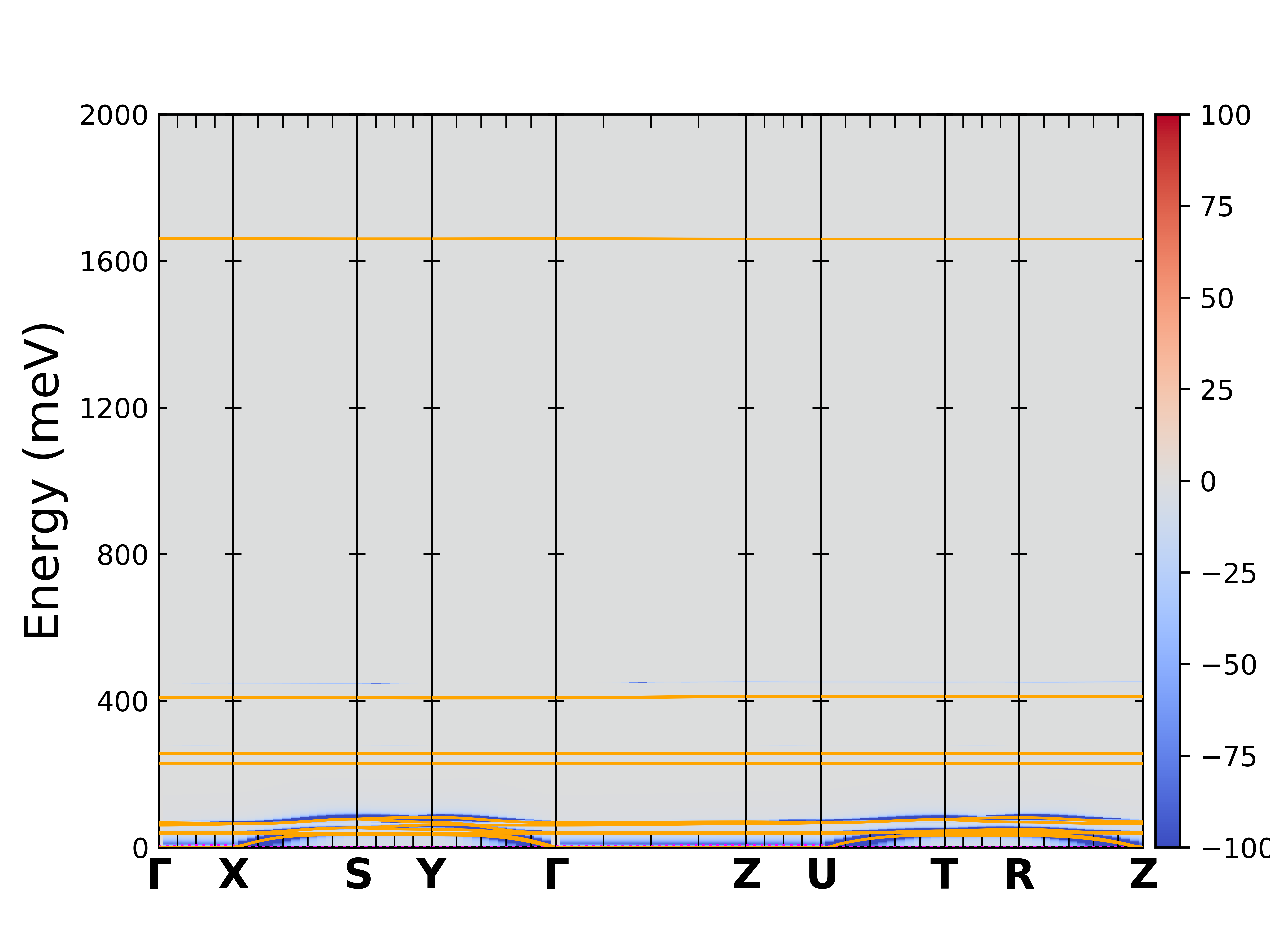}{}
    \caption{}
  \end{subfigure}
  \begin{subfigure}{8cm}\centering
    \includegraphics[width=8cm]{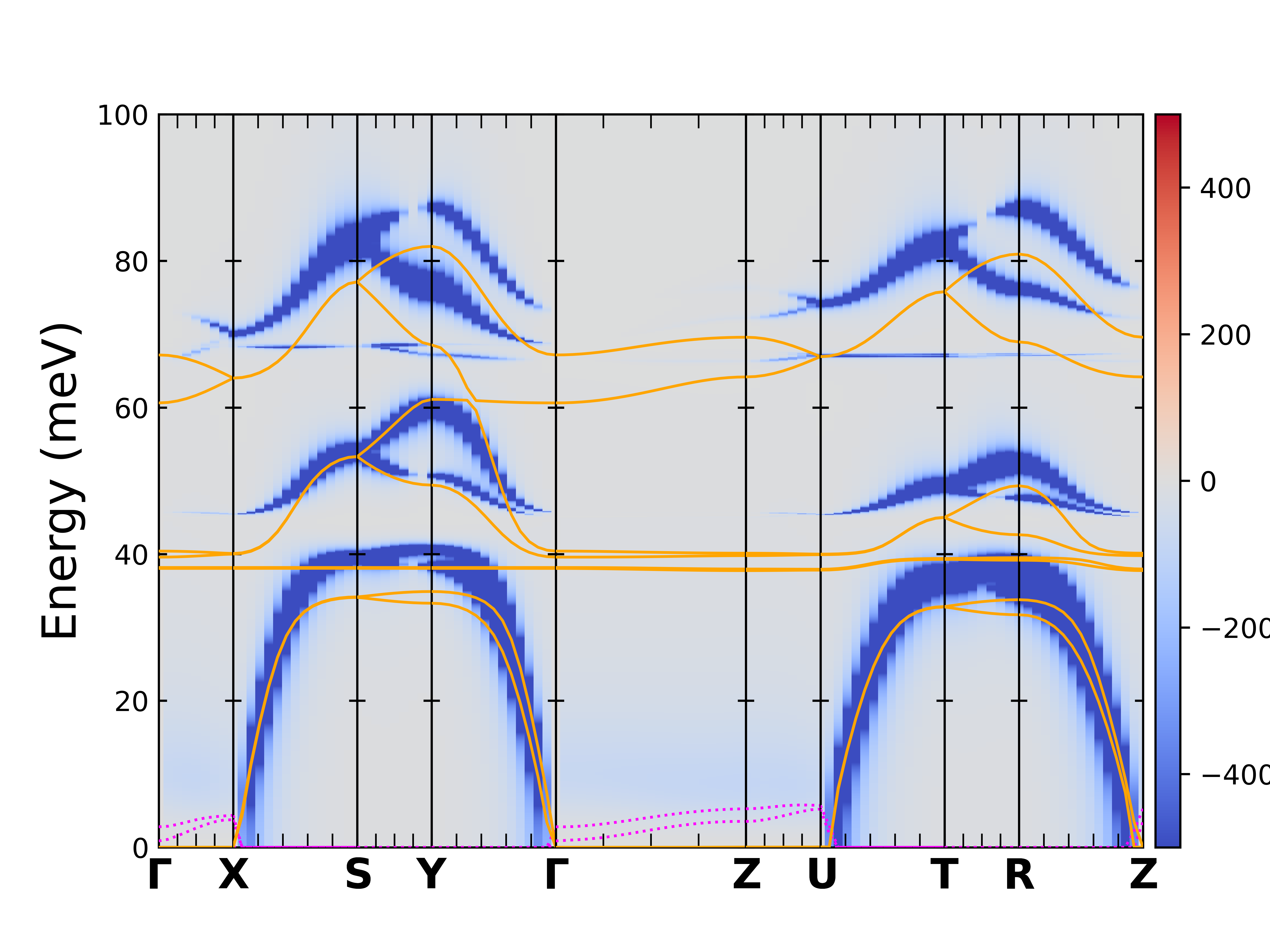}{}
    \caption{}
  \end{subfigure}  
  \caption{Spin wave dispersion and
    ${\rm Im}[{\rm Tr}[\chi^{+-}({\bf q},\omega)]]$ of LiV$_2$O$_5$ obtained from classical Heisenberg Hamiltonian including all sites: (a) $Pmmn$, (b)
  zoom in on energy lower region for $Pmmn$,  (c) $P2_1mn$ and (d) zoom in for $P2_1mn$ on low energy region. Yellow solid lines, spin wave energy, pink dotted line, imaginary eigenvalue,  ${\rm Im}[{\rm Tr}[\chi^{+-}({\bf q},\omega)]]$ is  indicated by color scale.}
\label{figspinw}
\end{figure*}

However, we have seen that exchange interactions with the bridge and vanadyl oxygen are among the largest inteactions in the system.
Thus we need to take this into account also for the spin waves.
As a first estimate we may note that the on-site terms in  the eigenvalue problem for the spin waves corresponding to O$_v$ are given by the term
$\sum_k J_{Ov,k}S_k/S_{O_v}$ and are dominated by the $k=V$ to which that O$_v$ is bonded. Again switching to the $J_{ij}$ with respect to spins, we see this diagonal term will be of order $\omega/M_{Ov}-4J_{Ov,V}/|M_{Ov}|^2$ and with $J_{Ov,V}\approx 14$  meV and $M_{Ov}=0.047$ we expect four  eigenvalues around
1200 meV.  Similarly, the $O_b$ terms include two interactions with V and with
$J_{Ob,V}\approx 6.8$ meV and $M_{Ob}=0.067$ we expect two eigenvalues dominated by O$_b$ in their eigenvector around 400 meV. In reality there are interactions between these diagonal terms and we find two eigenvalues near 690 meV and four near 1000 meV.

We can obtain a better approximation as follows. 
Including the two $O_b$ in the model of one ladder, we need to consider a $6\times6$ matrix. At this point it is useful to point out that the spin wave problem
of an antiferromagnet can be reduced to an eigenvalue problem for $\omega({\bf q})^2$ of a matrix of half the size because of the symmetry between up and down spin. The eigenvalues thus always come in pairs of opposite values.
This is shown in Appendix \ref{appa}.
Labeling the part of the matrix corresponding to the up spin V1,V3 and down  spin O$_{b1}$ as $J_{aa}$ and the matrix describing the interactions with their opposite spins along the  ${\bf b}$ chain as $J_{ab}$ we have
\begin{eqnarray}
  \bar{J}_{aa}&=&\left(\begin{array}{ccc}2J_1-J_2+J_b\frac{m_{O_b}}{m_V}&J_2&J_b\\J_2&2J_1-J_2+J_b\frac{m_{O_b}}{m_V}&J_b\\J_b&J_b&2J_b\frac{m_V}{m_{O_b}}\end{array}\right) \nonumber \\
       \bar{J}_{ab}&=&\left(\begin{array}{ccc}2J_1\cos{(q_yb/2)}&&\\&2J_1\cos{(q_yb/2)}&\\&&0
       \end{array}\right) \nonumber \\
\end{eqnarray}

With the Hamiltonian $H_{aa}=\bar{J}_{aa}\bar{J}_{aa}-\bar{J}_{ab}\bar{J}_{ba}$ and the 
``overlap matrix''
\begin{equation}
  B_{aa}=\left(\begin{array}{ccc}(1/m_V)^2&&\\&(1/m_V)^2&\\&&(1/m_{O_b})^2
  \end{array}\right)
\end{equation}
we then need to solve the generalized eigenvalue problem
$H_{aa'}X_a'=\omega({\bf q})^2B_{aa'}X_a'$, but since ${\bf B}$ is diagonal, this can be reduced to the standard eigenvalue problem for ${\bf B}^{-1/2}{\bf H}{\bf B}^{-1/2}$
or ${\bf M}{\bf H}{\bf M}$.
This matrix can be diagonalized analytically by Mathematica but  the resulting equations are a bit too cumbersome to give here. Inserting the values of the relevant moments and exchange interactions, we find one ``acoustic'' eigenvalue with the usual $|\sin{q_yb/2}|$ behavior and two almost flat modes at about about 150 meV and 674 meV. Inspecting the eigenvectors
shows that the mode at 150 meV has zero contribution from the O$_b$ and opposite sign for the spins on the two V in a rung. So, this corresponds to the optic mode we had before in the model with only the two V in a rung. In the high frequency mode centered about 674 meV, the V spins are parallel  but are also parallel to the O$_b$ which means the O$_b$ magnetic moment is opposite from that in the ground state, where they were antiparallel.   Finally, in the low frequency acoustic mode, the two V have the same sign spin but antiparallel to the magnetic moment of the O$_b$.

Similarly, let's consider the effect of the vanadyl oxygen interaction with the vanadium in the $ P2_1mn$ case where we need to consider only the large moment V and a single chain.
The matrices are then given by
\begin{eqnarray}
  J_{aa}&=&\left(\begin{array}{cc}0&J_v\\J_v&0
  \end{array}\right),\nonumber \\
  J_{ab}&=& \left(\begin{array}{cc}2J_1\cos{(q_yb/2)}
&0\\0&0
  \end{array}\right)\nonumber\\
  \bar{J}_{aa}&=&\left(\begin{array}{cc}-2J1+J_v\frac{m_{O_v}}{m_V}&J_v\\
    J_v&J_v\frac{m_V}{m_{O_v}}
  \end{array}\right)\nonumber \\
\end{eqnarray}
This gives again an acoustic mode with maximum of about 30 meV and a flat  optic mode near 1700 meV corresponding to spins on V precessing in phase  with the O$_v$ or in antiphase. Indeed in this case the highest modes corresponding to the $O_v$  bonded to the V$^{4+}$ gives a spin-wave mode at about 1660 meV.

At this point we have built enough understanding of the spin wave eigenvalues and we may look at the final result including all 32 atoms as magnetic.
The  spin wave spectra and the imaginary part of the trace of the $\chi^{+-}({\bf q},\omega)$ are  shown in Fig. \ref{figspinw} for LiV$_2$O$_5$ $Pmmn$
as calculated within the linear spin wave theory including all atoms (regardless of the size of their magnetic moment) in the classical Heisenberg model.  Besides the lower two  spin waves (shown in the zoom in  Fig. \ref{figspinw}b) which behave linear near $\Gamma$ as required for an antiferromagnet, several features are noteworthy.
Most of all, we see several almost dispersionless modes, four with energy near 1010 meV, four near 830 meV, two near 690 meV, two with somewhat larger dispersion near 240 meV, two near 103 meV, besides the lower two which have a maximum near 50 meV. In total there are 16 modes with positive frequency and another 16 with equal magnitude but opposite sign because we include all 32 atoms in the cell.
As explained above,  these modes have eigenvectors dominated by O$_v$, O$_c$, O$_b$, V, Li, V respectively from high to low energy.
For example, the Li has interactions with the two V lying in the same $m_y$ mirror plane which predict a diagonal term at $4J_{\rm Li,V}/m_{\rm Li}=4\times0.0033/0.000146= 90 $ meV and indeed we find two eigenvalues related to Li near 106 meV.

We also note that the lowest mode becomes imaginary (indicated by pink dotted line) along the $\Gamma Z U$ lines and even the second mode becomes imaginary halfway along $\Gamma Z$.
This is caused by the antiferromagnetic exchange interactions between layers and indicates that the reference system is not the lowest energy configuration.
It predicts that spins would alter up and down along the layers in the ${\bf c}$ direction.

The flat modes can be viewed as an analog of optic modes in phonon spectra and correspond to modes in which the spins on the different atoms precess independently  of each other instead of staying locked in phase as in the ground state. The heat map indicates the
${\rm Im}[{\rm Tr}[\chi^{+-}({\bf q},\omega)]]$. We can see that for the lower modes it follows the spin wave although the intensity is weaker along $\Gamma X$ and $\Gamma Z$. For the higher  modes, we can see that it is shifted from the eigenvalues. This is because the latter are obtained from the bare $\chi^{0+-}$ while the heatmap is obtained from the dressed dynamic $\chi^{+-}$ and the peaks indicate the poles of the latter. We may also notice that the susceptibility
peaks only in certain regions of {\bf q}-space and is invisibly small for the modes near 106 meV associated with Li and near 830 meV associated with O$_c$.
These  are the atoms with smallest magnetic moments and also small exchange interactions.

In part (c) and (d) of Fig. \ref{figspinw} we can see that in $P2_1mn$ we have even higher frequency modes  which is related to the very  strong $J_v$ interaction but because of the smaller exchange between the small and large moment vanadiums and also smaller interaction with the bridge oxygen the other optic modes are more compressed together in the low energy region up to about 400 meV.
The strong mixing of modes in the range 0-100 meV is apparent. 

In summary of our analysis of the spin waves obtained from our distributed moment model, we obtain a very complex spectrum with various optical modes related to the independent precession of the vary small moments. However, this seems counterintuitive and thus we have to question whether such modes correspond to a valid new kind of excitations of the spin system or are an artifact of the model or treating these small moments as magnetic degrees of freedom. For example, it seems rather strange  that the modes relating to the very small induced moments on the Li atom precessing out of sync with the V spins would significantly interfer with the lowest spin wave as seen in Fig. \ref{figspinw}d. 
To end this section, we thus need to critically examine the validity of treating the Heisenberg model with independent degrees of freedom on all atoms for the spin dynamics. In fact, we will argue that these high energy modes violate the adiabatic approximation and are thus not valid excitations of the system. 

Our results here on the spin waves are obtained within linear spin wave theory (LSWT) and using a classical Hamiltonian with unit vectors allowing the spins to rotate in any direction. However, in the $P2_1mn$ case we actually have a $S=1/2$  linear chain.
The spin waves in LSWT are obtained usually from the Holstein-Primakoff transformation and are valid in the large $S$  limit and correspond to the first term in a $1/S$ expansion. For small spins, in particular for $S=1/2$ higher order terms corresponding to magnon-magnon interactions are expected to be important.
For the $Pmmn$ case, the classical LSWT model is even more problematic because it would correspond to an unphysical $S=1/4$. As we mentioned before, this results from the way we treat the magnetic moments as localized on atoms while in the present case, a spin $S=1/2$ is really distributed over several atoms in a molecular type state, which even includes important contributions in the vanadyl and bridge oxygens. The meaning of the spin wave then becomes less clear.
The optic spin wave here   would correspond to the spins precessing opposite on the two atoms of the rung and more generally a complex patterns of how the spins on the V and O precess relative to each other. This seems  impossible if it corresponds really to a localized single electron spin per rung.   On the other hand, if we view the  classical
Heisenberg model unit vectors as representing the local direction at each point in space  of a continuous spin density
as we do in spin density functional theory, which we discretized or course-grained by assigning a single moment per muffin-tin sphere, then these unusual spin waves would represent a more complex excitation of the magnetic system, not captured by the picture of local spins.

The dynamic susceptibility itself is a well defined correlation function and has physical meaning beyond the Heisenberg model with static  exchange interaction on which we map the problem.  Thus, the fact that the peaks in the imaginary part of the susceptibility track the spin-wave eigenvalues lends validity to this approach even for a more itinerant spin case where the spin density is distributed over several atoms and is not an integer or half integer spin per atom.

At present there are no experimental data available supporting such unusual spin wave type excitation but to the best of our knowledge, nobody has looked for them. On the other hand, for the high energy modes predicted here, the energy becomes comparable to the spin-up to spin-down energy differences, in other words to the longitudinal energy scale. This means that the adiabatic decoupling of longitudinal and transverse excitations underlying the construction of the
Heisenberg Hamiltonian itself may break down and invalidate the existence of these high energy modes.
Even for the low energy modes related to small moment sites, like Li or Na or O$_c$ but which have also low exchange interactions, the existence of these modes
is questionable. In fact, we can estimate a local longitudinal energy scale by locally quenching the magnetic moment adding an external field inside the sphere in a constrained moment approach. These constrained moment calculations give the magnetic field $\delta B$ along the spin direction that is needed to quench the moment. The associated energy cost is
$\delta E=m\delta B +\frac{1}{2} \chi(\delta B)^2$ with $\chi$ the local longitudinal susceptibility. For the O$_v$ and O$_b$ these energies are of order 20-30 meV. They are expected to be significantly smaller for the even smaller moment sites O$_c$ and Li or Na. It is not entirely obvious what to compare them to but if we compare them to the corresponding spin wave energies that are dominated by these small moment atoms, then the longitudinal energy scale is in fact smaller than instead of larger than the transverse energy scale and this would indicate that the adiabatic decoupling of transverse from longitudinal excitations is no longer valid. This means we cannot treat these small moments as independent dynamical degrees of freedom. 
We conclude that including the small moment  exchange interactions
is fine for the static properties and low energy spin wave excitations which determine the critical temperature but leads to unphysical optical type spin waves related to independent precession of the small moments.
In fact, quenching the moments on these oxygen atoms costs less energy than a transverse excitation in which the O moments precesses opposite to its antiferromagnetic alignment with V moments and the latter are thus not a valid excitation. 

This reasoning cannot directly be used to dismiss the spin waves in which the two V precess opposite to each other. However, we need to compare the energy scale of  the transverse mode in which V precess opposite to each other with that of a longitudinal excitation in which the magnitude of the spin on the V is changing. The $Pmmn$ structure should in fact be viewed as a time average of the spin fluctuating between two adjacent V across a bridge. Since the $Pmmn$ structure has an energy of about 0.1 eV/atom higher than the $P2_1mn$ structure, we can see that the longitudinal energy scale is in fact lower than the optical spin wave excitation. Furthermore, if the spins are fluctuating in a disorderly manner between the two V sites, a coherent spin wave excitation would no longer be possible.

We  thus conclude that only the lowest energy spin wave branches, in which the moments  stay locked together on the different atoms as in their ground state, are valid. The overall spin $S=1/2$ stays  a rigid spin  during its transverse spin-wave excitation. In other words, its shape or distribution over the different sites stays fixed. In fact, this strong locking of the spins is enforced by the strong exchange interactions $J_2$ keeping the two V spins paralle and the even stronger $J_v$ and $J_b$ keeping the small induced moment on the vaxnadyl and bridge oxygens antiparallel to the moment on the V.

The available inelastic neutron scattering measurements are restricted to a small energy range up to about 40 meV and apply to the low temperature phase of NaV$_2$O$_5$ below the
charge disproportionation transition of 34 K 
\cite{Grenier2001,Yoshihama1998}. They were discussed in \cite{Lambrechtnav2o5}.
The maximum of the lowest spin wave along $\Gamma Y$ in those measurements show significant uncertainty but is consistent with our value of 50 meV and could be interpreted as a 1D  effective $S=1/2$ Heisenberg model along the chain
with a dispersion $\omega_{\bf q}=|2J_{eff}\sin{(q_yb/2}$ with 
an effective $J_{eff}$ of about $-25$  meV which in turn is consistent with Isobe and Ueda's initial estimate from the maximum of the susceptibility \cite{Isobe1996}.

\section{Conclusions}
In this paper, we studied the electronic band structure, optical properties and magnetic exchange interactions in two crystal structure  of LiV$_2$O$_5$ and  NaV$_2$O$_5$, one ($P2_1mn$) showing charge disporportionation into $V^{4+}$ and V$^{5+}$ and another one ($Pmmn$) in which the two V are equivalent and nominally V$^{4.5+}$.
We find small differences in the band structures, obtained using the QS$GW$ method, with slightly larger gap between occupied $d_{xy}$  and lowest empty $d_{xy}$ states in $P2_1mn$ of LiV$_2$O$_5$.  This applies both to the ferromagnetic and the lower energy antiferromagnetic arrangement of the spins in a cell doubled along the ${\bf b}$ direction.
In NaV$_2$O$_5$, we find  similar bands but with somewhat larger dispersion of the lowest empty bands indicating additional hybridization with the $d_{xz}$, $d_{yz}$ orbitals. 
In terms of the optical dielectric function, we find excitonic transitions just below 2 eV
related to the $d-d$ transitions, with a notable anisotropy between the $a$ and $b$ polarizations which becomes even stronger in the $P2_1mn$ structure.

The magnetic moments
show that the charge disproportionation indeed occurs in the $P2_1mn$ case and is driven by the polaronic distortion around the V$^{4+}$ that carries the magnetic moment and pushes nearby O away from it and attracts the Li closer to it. In reality the charge disproportionation is not perfect and we still find a small moment of about 0.1 $\mu_B$ on the V$^{5+}$. We find a somewhat less pronounced charge disproportionation for NaV$_2$O$_5$.
Furthermore, we note  significant moments, about one order of magnitude smaller induced on the vanadyl and bridge oxygens but which play a key role in the exchange interactions. 

To study the magnetic exchange interactions, we adopt a distributed moment approach in which exchange interactions are calculated between moments on sites, including both  vanadium  and oxygen
sites even though these magnetic moments correspond to a single $S=1/2$  in an extended  molecular type orbital corresponding to the narrow isolated band which becomes spin-polarized.
The exchange interactions were derived from a linear response approach and pertain to a classical Heisenberg type model extracted from the transverse spin susceptibility calculated from the QS$GW$ band structures following the approach of Kotani and van Schilfgaarde\cite{Kotani2008}. This approach uses an integration over spheres weighting the susceptibility by a projector proportional to the spin density inside the sphere.

The exchange interactions in the $Pmmn$ case show a strong ferromagnetic coupling between the two V on a rung, which may seem obvious if we view it as a local spin distributed over two V. They show antiferromagnetic coupling between V along nearest neighbors along the chain direction. Other V-V interactions are smaller and show some degree of geometric frustration. Interactions between layers are smaller but antiferromagnetic indicating that spins might order antiferromagnetically from layer to layer. Most surprisingly, we find very strong exchange interactions between the small moments induced on vanadyl and bridge oxygens with the moments on their neighboring vanadium atoms. These are in fact the strongest exchange interactions in the system in this site-by-site  analysis of the exchange interactions. In the $P2_1mn$ case, the interactions between small moment nominally V$^{5+}$ atoms and between large and small moment V are strongly reduced and support a more isolated chain picture.
However, the exchange interactions between V in the chain and with the vanadyl oxygens in that chain are then more than twice as large. The exchange interactions in the Na and Li case are of  comparable magnitude. 

The N\'eel temperatures calculated in the mean field approximation based on these exchange interactions are calculated.
The role of the major exchange interaction parameters was investigated by constructing analytical models. The inclusion of the vanadium-oxygen exchange interactions is found to be crucial. Without it, the critical temperature would be underestimated by a factor two. The classical Heisenberg model gives temperatures of 187 K  in $Pmmn$ and 293 K in the $P2_1mn$ case. Because ultimately, our distributed moment description corresponds to an overall  $S=1/2$ system, a quantum correction factor $(S+1)/S=3$ is included and then gives mean field temperatures comparable with or larger than the experimental one. Applying a typical reduction factor of 0.56 for effects beyond mean field brings the estimate N\'eel temparature close to the experimental values for both the $Pmmn$ and $P2_1mn$ structures. This is further substantiated by an RPA calculation which reduces the critical temperature by a factor 0.8-0.9 compared to the MFA for the $Pmmn$ case.
The RPA estimate for $Pmmn$ is in reasonable agreement with experiment for the case of NaV$_2$O$_5$. As we find here somewhat larger exchange interactions for the Li case, we predict a somewhat higher critical temperature. 

Our distributed moment model leads to very complex spin wave spectra.
Besides the usual acoustic branch the distributed moment classical Heisenberg model leads to several high energy optical branches with little dispersion and which correspond to more complex spin excitations in which the moments on the different sites precess in such a way that they change the relative spin orientations found in the reference state. For example the spins on the two V in a rung in some of these modes
are then precessing opposite to each other instead of in phase with each other.
While these modes can be understood in detail within the model, the applicability of the Heisenberg model for these high energy excitations itself becomes
questionable. Within a strict rigid spin approximation, in which the spin, however distributed its probability density is over several atoms, is required to precess as a rigid entity, such excitations are not allowed. This inability of the distributed moment model to properly describe the spin dynamics, however, does not invalidate its usefulness to calculate the energy scale of the effective exchange interaction between the true spins, which determines the critical temperarture and the dispersion of the lowest energy spin waves in which the magnetic moments in the differenet spheres stay coherently in sync with each other.

Our distributed moment approach presents a complementary point of view to the prevalent local spin in rung of a quarter filled ladder compound that has been used in most of the prior literature and tests to what extent a simple Heisenberg model with a single antiferromagnetic exchange coupling is valid.

\acknowledgements{This work was supported by the U.S. Department of Energy Basic Energy Sciences (DOE-BES) under Grant No. DE-SC0008933. Calculations made use of the High-Performance Computing Resource in the Core Facility for Advanced Research Computing at Case Western Reserve University, the Ohio Supercomputer Center, and the National Energy Research Scientific Computing Center, a DOE Office of Science user facility supported under Award No. DE-AC02-05CH11231 using NERSC Award No. BES- ERCAP0021783. C.G.'s stay at NREL  (now NLR) was supported by the Science Graduate Student Research Program. S.A. and M.v.S are supported by the Computational Chemical Sciences program within the U.S. DOE, Office of Science, BES, under Award No. DE-AC36-08GO28308. J.J. acknowledges support under the CCP9 project Computational Electronic Structure of Condensed Matter [part of the Computational Science Centre for Research Communities (CoSeC)}   

{\bf Data Availability}
The data that support the findings of this article are openly available\cite{DATA}

\appendix
\section{Reduction to $\omega^2$ eigenvalue problem.} \label{appa}
In this appendix we show that the spin-wave eigenvalue problem can be related to an eigenvalue problem for $\omega^2$ and hence all eigenvalues come in equal magnitude positive and negative pairs.  We then apply it to the case of $Pmmn$ when keeping only the main interactions and find analytical expressions for the spin waves. We write the exchange interaction matrix $\bar{J}$ from
Eq.(\ref{eq:spwv}) in block form where all up spins correspond to $a$ and down spins to $b$ and the atoms in each subblock are ordered the same way so that the translation symmetry by ${\bf b}$ which relates up and down spin atoms applies.
The spin wave eigenvalue problem then has the form
\begin{equation}
  \left(\begin{array}{cc}\bar{J}_{aa}({\bf q})-\frac{\omega}{|m_a|}\delta_{aa}& \bar{J}_{ab}({\bf q})\\
      \bar{J}_{ba}({\bf q})& \bar{J}_{bb}({\bf q})+\frac{\omega}{|m_b|\delta_{bb}}
  \end{array}\right)\left(\begin{array}{c}C_a\\C_b
  \end{array}\right)=\left(\begin{array}{c}0\\0
  \end{array}\right),
\end{equation} 
We now use that $\bar{J}_{aa}({\bf q})=\bar{J}_{bb}({\bf q})$ and $m_b=-m_a$.
Writing these out block by block and solving the second equaiton for $C_b$ and inserting in the first equation yields
\begin {equation}
  \left(\bar{J}_{aa}-\frac{\omega}{|m_a|}\right)C_a-\bar{J}_{ab}\left(\bar{J}_{aa}+\frac{\omega}{|m_a|}\right)^{-1}\bar{J}_{ba}C_a=0.
\end {equation}
If the matrices $\bar{J}_{aa}({\bf q})$ and $\bar{J}_{ab}({\bf q})$ commute then this leads to
\begin{equation}
  \left(\bar{J}_{aa}({\bf q})\bar{J}_{aa}({\bf q})-\bar{J}_{ab}({\bf q})\bar{J}_{ba}({\bf q})-\frac{\omega^2}{|m_a|^2}\right)C_a=0, \label{eq:Jeff}
\end{equation}
which is a generalized eigenvalue problem for $\omega^2$ with a  diagonal and positive definite overlap matrix $1/|m_a|^2$ of half the matrix size of the original problem.
If all spin up magnetic moments are equal, as is the case for $Pmmn$ and considering only V spins, then
the matrix $1/|m_a|^2$ is proportional to the unit matrix and we just have a
normal eigenvalue problem  with eigenvalues $\lambda=\omega^2/|m|^2$ and obviously all spin wave frequencies $\omega_j$ come in  pairs
$\omega_j({\bf q})=\pm |m|\sqrt{\lambda_j({\bf q})}$.
The proof of the commutation of the two matrices rests on the symmetries of these matrices.
\section{Application to $Pmmn$ including V only.} \label{appb}
In the $Pmmn$ case we have a $8\times8$ matrix and both the $\bar{J}_{aa}$ and $\bar{J}_{ab}$ matrices are of the form
\begin{equation}
  \left(\begin{array}{cccc}Af({\bf q})&Bg({\bf q})&Ch({\bf q})&Dj({\bf q})\\Bg({\bf q})^*&Af({\bf q})&Dj({\bf q})&Ch({\bf q})\\Ch({\bf q})^*&Dj({\bf q})^*&Af({\bf q})&Bg({\bf q})\\Dj({\bf q})^*&Ch({\bf q})^*&Bg({\bf q})^*&Af({\bf q})
  \end{array}\right),\label{eq:matrixform}
\end{equation}
where $f({\bf q}),g({\bf q}),h({\bf q}),j({\bf q})$ are sums over phase factors of the type $\sum_{{\bf T}}e^{i{\bf q}\cdot{\bf T}}$. For $\bar{J}_{aa}$,
  if we include only the interactions $J_1\dots J_6$, $f=g=j=1$ and
  $h({\bf q})=(1+e^{-iq_xa})$ while for $\bar{J}_{ab}$,
  $f({\bf q})=(1+e^{-iq_yb})$, $g=j=1$ and $h({\bf q})$ is the same as for
  $\bar{J}_{aa}$. Writing $h({\bf q})=(1+\cos{q_xa})-i\sin{q_xa})=h'-ih''$
  the matrices can be written 
\begin {eqnarray}
 \bar{J}_{aa}&=&A_{aa}\sigma_0\otimes\tau_0+B_{aa}\sigma_x\otimes\tau_0+C_{aa}h'\sigma_0\otimes\tau_x\nonumber \\
 &&+C_{aa}h''\sigma_0\tau_y+D_{aa}\sigma_x\otimes\tau_x,\nonumber\\
  \bar{J}_{ab}&=&A_{ab}\sigma_0\otimes\tau_0+B_{ab}\sigma_x\otimes\tau_0+C_{ab}h'\sigma_0\otimes\tau_x\nonumber \\
 &&+C_{ab}h''\sigma_0\tau_y+D_{ab}\sigma_x\otimes\tau_x,\nonumber\\
\end {eqnarray}
where $\sigma_x$ and $\tau_x$, $\tau_y$  are Pauli matrices and $\sigma_0$, $\tau_0$ are $2\times2$ unit matrices, with the $\tau$ matrices describing the outer structure of the matrix in terms of $2\times2$ blocks and the $\sigma$ martrices describing the $2\times2$ inner blocks and $\otimes$ denotes the Kronecker outer product. 
This form results simply from the symmetry relations between the atoms 1-4 in the unit cell for all spin up and 5-6 for all spin down sites.
Because of this form, and the commutation relations of the Pauli $\sigma_x$ and unit matrices, the commutation of $\bar{J}_{aa}$ and $\bar{J}_{ab}$ are indeed satisfied and the reduction to a $\omega^2$ eigenvalue problem is satisfied.
The only terms involving $[\tau_x,\tau_y]=2i\tau_z$ which do not commute occur in pairs  that cancel each other because the same sum of phase factors occurs
in $C_{aa}h({\bf q})$ and $C_{ab}h({\bf q})$, \ie we have
$C_{aa}C_{ab}h'h''\sigma_0\otimes([\tau_x,\tau_y]+[\tau_y,\tau_x])$.
The effective  exchange matrix of Eq.(\ref{eq:Jeff}) is then of the same form
without the phase factors which now all become real and can be absorbed in the $A\dots D$ and has eigenvalues $A+B+C+D$, $A+B-C-D$, $A-B-C+D$, $A-B+C-D$. Thus all eigenvalues can be obtained fully analytically.

For example including the interactions $J_1\dots J_6$ of Table \ref{tabexchange}
the matrices $\bar{J}_{aa}$ and $\bar{J}_{ab}$ have the following form
$\bar{J}_{aa}({\bf q})=(2J_1-J_2-J_3+J_4 -2J_5+2J_6)\sigma_0\otimes\tau_0 +J_2\sigma_x\otimes\tau_0+J_3\sigma_x\otimes\tau_x+J'_5({\bf q})\sigma_0\otimes\tau_x+J''_5({\bf q})\sigma_0\otimes\tau_y$
with $J_5({\bf q})=J_5(1+e^{-iq_xa})$ and $J'_5$ and $J''_5$ its real and imaginary parts. Similarly 
$\bar{J}_{ab}({\bf q})=J_1({\bf q})\sigma_0\otimes\tau_0+J_4\sigma_x\otimes\tau_x+J'_6({\bf q})\sigma_0\otimes\tau_x+J''_6({\bf q})\sigma_0\otimes\tau_y$ with $J_1({\bf q})=J_1(1+e^{-iq_yb})$, $J_6({\bf q}=J_6(1+e^{-iq_xa})$.
Working out the matrix products, we then finally obtain the matrix to be diagonalized is of the form of Eq.(\ref{eq:matrixform}) with $f=g=h=j=1$ and 
\begin{eqnarray}
  A&=&(2J_1-J_2-J_3+J_4-2J_5+2J_6)^2+J_2^2+J_3^2\nonumber \\
  &&+2J_5^2(1+cos{q_xa})-2J_1^2(1+\cos{q_yb})\nonumber\\&&-J_4^2-2J_6^2(1+\cos{q_xa}),\nonumber \\
  B&=&2J_2(2J_1-J_2-J_3+J_4-2J_5+2J_6)\nonumber \\&&+2J_3J_5(1+\cos{q_xa})\nonumber \\
  &&-2J_4J_6(1+\cos{q_xa}),\nonumber \\
  C&=&2J_1J_6(1+\cos{q_xa}+\cos{q_yb}+\cos{(q_xa-q_yb)})+2J_2J_3\nonumber \\
  &&+2J_5(1+\cos{q_xa})(2J_1-J_2-J_3+J_4-2J_5+2J_6),\nonumber \\
  D&=&2J_3(2J_1-J_2-J_3+J_4-2J_5+2J_6)\nonumber \\
  &&2J_2J_5(1+\cos{q_xa})-2J_4J_1(1+\cos{q_yb})\nonumber \\
\end{eqnarray}

If we only keep the interactions $J_1\dots J_4$ there is no $q_x$ dispersion
and the problem further simplifies.
\begin{eqnarray}
  \omega_1^2({\bf q})&=&4|m|^2J_1(J_1+J_4)\sin^2{(q_yb/2)},\nonumber \\
  \omega_2^2({\bf q})&=&4|m|^2[J_1(J_1-J_4)\sin^2{(q_yb/2)}+J_3^2-2J_1J_3-J_3J_4\nonumber\\&&+2J_1J_4],\nonumber \\
  \omega_3^2({\bf q})&=&4|m|^2[J_1(J_1+J_4)\sin^2{(q_yb/2)}+J_2(J_2-2J_1)],\nonumber \\
  \omega_4^2({\bf q})&=&4|m|^2[J_1(J_1-J_4)\sin^2{(q_yb/2)}+J_3^2+J_2(J_2-2J_1)\nonumber \\
    &&+2J_2J_3+2J_1J_4-2J_1J_3-J_3J_4].\nonumber\\
\end{eqnarray}
When $J_3=J_4=0$ these agree with two  doubly degenerate eigenvalues of uncoupled ladders as given in 
Eq.(\ref{eq:sw1}).
\newpage
\bibliography{Bib/linav2o5,Bib/lmto,Bib/gw,Bib/dft,Bib/spinchi}
\newpage
\section{Supplemental information : Charge disproportionation and spin ordering in LiV$_2$O$_5$ and NaV$_2$O$_5$: a distributed moment point of view.}

\subsection{Structural parameters}
  The structural parameters used here were either obtained from the Materials Project (MP) \cite{MPcitation} or from experimental data in the form of .cif files from the International Crystallographic Structure Database(ICSD).  
  For NaV$_2$O$_5$ we used the experimental data for $P2_1mn$ from \cite{Carpy1975,Galy1966}, ICSD entry 1898.
  For the $Pmmn$ structure we used the data from \cite{Meetsma1998} corresponding to ICSD  entry  59345.
  For LiV$_2$O$_5$ we used the MP data from mp-18934 for $Pmmn$ and
  mp-555869 for $P2_1mn$. The setting of the latter spacegroup there is $Pmn2_1$ meaning that the $a$ axis is the chain direction, $b$-axis is perpendicular to the layers and the c-axis is in the plane perpendicular to the chains and along the rungs of the ladder. For easier comparison, we converted  the Brillouin zone to the $P2_1mn$ setting with the crystallographic axes of the two structures coinciding.

  The lattice constants are summarized in Table \ref{tabstruc}.

  \begin{table}[t]
    \caption{Lattice constants used in \AA.\label{tabstruc}}
    \begin{ruledtabular}
      \begin{tabular}{ccc|cc}
          &\multicolumn{2}{c|}{LiV$_2$O$_5$}& \multicolumn{2}{c}{NaV$_2$O$_5$}\\\hline
              &$Pmmn$ & $P2_1mn$&$Pmmn$ & $P2_1mn$ \\ \hline
              a& 11.488 & 11.546 & 11.311 & 11.329 \\
              b& 3.622  & 3.589 & 3.610 & 3.612 \\
              c& 4.727 & 4.854 & 4.800 & 4.791 \\
      \end{tabular}
              \end{ruledtabular}
  \end{table}
   \begin{figure*}
     \begin{subfigure}{8cm}\centering
       \includegraphics[width=8cm]{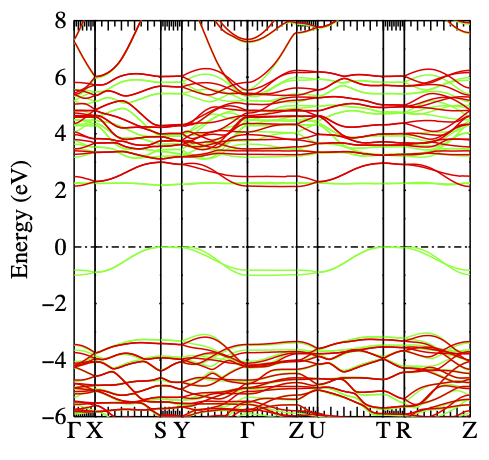}{}
       \caption{}
     \end{subfigure}
      \begin{subfigure}{8cm}\centering
        \includegraphics[width=8cm]{bnds-nav2o5-p21mn-fm.png}{}
          \caption{}
      \end{subfigure}
       \begin{subfigure}{8cm}\centering
         \includegraphics[width=8cm]{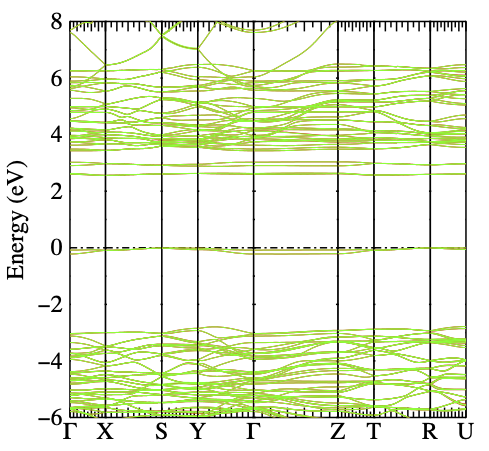}{}
          \caption{}
       \end{subfigure}
       \begin{subfigure}{8cm}\centering
                  \includegraphics[width=8cm]{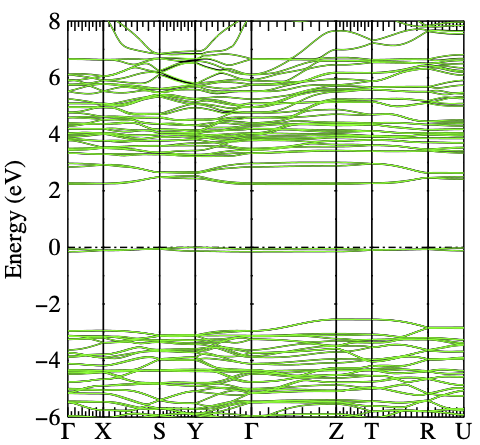}{}
           \caption{}
       \end{subfigure}
   \caption{Energy band structures of NaV$_2$O$_5$ in (a) $Pmmn$ FM, (b),$P2_1mn$ FM, (c) $Pmmn$ AFM and (d) $P2_1mn$ AFM cases.\label{nav2o5bands}}
 \end{figure*}

  \section {Molecular model for $d_{xy}$ band splitting.}
  Focusing on the two V-$d_{xy}$ orbitals with energy $\epsilon_d$ and the O-$p_y$ orbital the Hamiltonian has the form
  \begin {equation}
    H=\left(\begin{array}{ccc} \epsilon_d&&V_1\\
      &\epsilon_d&V_2\\
      V_1&V_2&\epsilon_p
    \end{array}
\right)
  \end {equation}
 
  Assuming $|\epsilon_d-\epsilon_p|\gg V_i$, $i=1,2$, in second order perturbation (or L\"owdin downfolding), we
  have an effective Hamiltonian 
  \begin{equation}
    \tilde{H}_{dd}=\left(\begin{array}{cc}\epsilon_d+U_1&U_{12}\\U_{12}&\epsilon_d+U_2\\
    \end{array}\right)
  \end{equation}
  with $U_1=V_1^2/(\epsilon_d-\epsilon_p)$,  $U_2=V_2^2/(\epsilon_d-\epsilon_p)$
  and  $U_{12}=V_1V_2/(\epsilon_d-\epsilon_p)$. $\tilde{H}_{dd}$ has eigenvalues
  \begin{eqnarray}
   \epsilon_\pm&=&\epsilon_d+\frac{U_1+U_2}{2}\pm\sqrt{\left(\frac{U_1-U_2}{2}\right)^2+U_{12}^2} \nonumber \\
   &=&\epsilon_d+ \frac{V_1^2+V_2^2}{2(\epsilon_d-\epsilon_p)}\pm\frac{V_1^2+V_2^2}{2(\epsilon_d-\epsilon_p)}\nonumber \\
   &=&\epsilon_d,\epsilon_d+\frac{V_1^2+V_2^2}{(\epsilon_d-\epsilon_p)} 
  \end{eqnarray}
  In $Pmmn$, $V_1=V_2=V$ and the splitting becomes $2V^2/(\epsilon_d-\epsilon_p)$, while in $P2_1mn$, $V_1=V-\delta V$, $V_2=V+\delta V$, so the splitting becomes $[2V^2+2(\delta V)^2]/(\epsilon_d-\epsilon_p)$. Thus the splitting is minimal for $\delta V=0$ but the difference is only of second order in $\delta V$.
  We may view $2V^2/(\epsilon_d-\epsilon_p)$  as a covalent  and $2(\delta V)^2/(\epsilon_d-\epsilon_p)$ as an ionic contribution to the gap.

\subsection{Energy bands for NaV$_2$O$_5$}

\begin{figure*}[t]
  \begin{subfigure}{7cm}\centering
    \includegraphics[width=7cm]{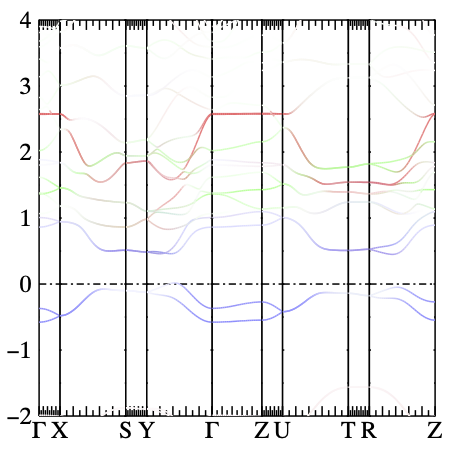}{}
    \caption{} 
  \end{subfigure}
  \begin{subfigure}{7cm}\centering
    \includegraphics[width=7cm]{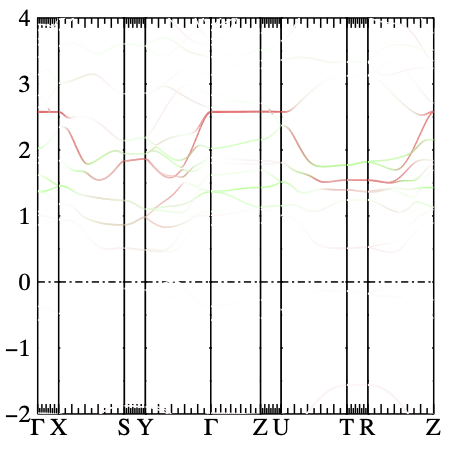}{}
    \caption{}
  \end{subfigure}
  \caption{Energy bands of FM $P2_1mn$ NaV$_2$O$_5$ indicating orbital contribiutions as blue $xy$, red $yz$, green $xz$ on the V$^{4+}$ atoms, (a) with and (b) without the main $xy$ contribution.\label{figorbband}}
\end{figure*}

The energy bands of NaV$_2$O$_5$ in FM and AFM $Pmmn$ and $P2_1mn$ are given in Fig.\ref{nav2o5bands}.
We may note that the lowest empty bands here show larger dispersion and the gap between occupied and empty
$d$-bands is somewhat decreased. Fig. \ref{figorbband} shows that these bands
have mainly $xy$ character but also a slight $yz$ component in particular near $Y$ and this contribution appears to be related to the downward bending of the first empty band near $Y$. This increased $yz$  contribution is in turn related to the distance from the V$^{4+}$ to the chain oxygens being increased and by sharpening the angle between O$_c$-V-O$_c$ along the $b$-chain.  This reduces the V-$d_{yz}$-O$_c$-$p_y$ antibonding interaction, lowering the $yz$ orbital and hence letting it participating more in the lowest empty  band.  Thus, subtle distortions of the structure related to the polaronic self-trapping of the electron near the V$^{4+}$
are responsible for these changes in the band dispersion. 
\subsection{Exchange interactions and spin-waves in NaV$_2$O$_5$}

\begin{table}
   \caption{Exchange interactions in NaV$_2$O$_5$ in meV  as defined with respect to the classical Heisenberg Hamiltonian with unit vectors. $N_e$ is the number of equivalent interactions.} \label{tabexchange-na}
   \begin{ruledtabular}
     \begin{tabular}{cccc}
       $J_n$&$N_e$&\multicolumn{1}{c}{$Pmmn$}&\multicolumn{1}{c}{$P2_1mn$} \\ \hline
       $J_1$&2& -2.99& -3.88\\
       $J_1^\prime$&2& & -0.73 \\
       $J_2$&1& 6.53 & 3.68 \\
       $J_3$&1& -0.90 & -0.68 \\
       $J_4$&1& -0.96 & -0.70 \\
       $J_5$&2& 0.24  & 0.17 \\
       $J_6$&2& 0.23  & 0.17 \\
       $J_7$&2& -1.16 & -0.08 \\
       $J_v$&1& -11.89 & -22.74 \\
       $J_b$&1& -7.06 & -2.93 \\
     \end{tabular}
   \end{ruledtabular}
\end{table}
  
\begin{figure*}[h]
  \begin{subfigure}{5cm}\centering
    \includegraphics[width=5cm]{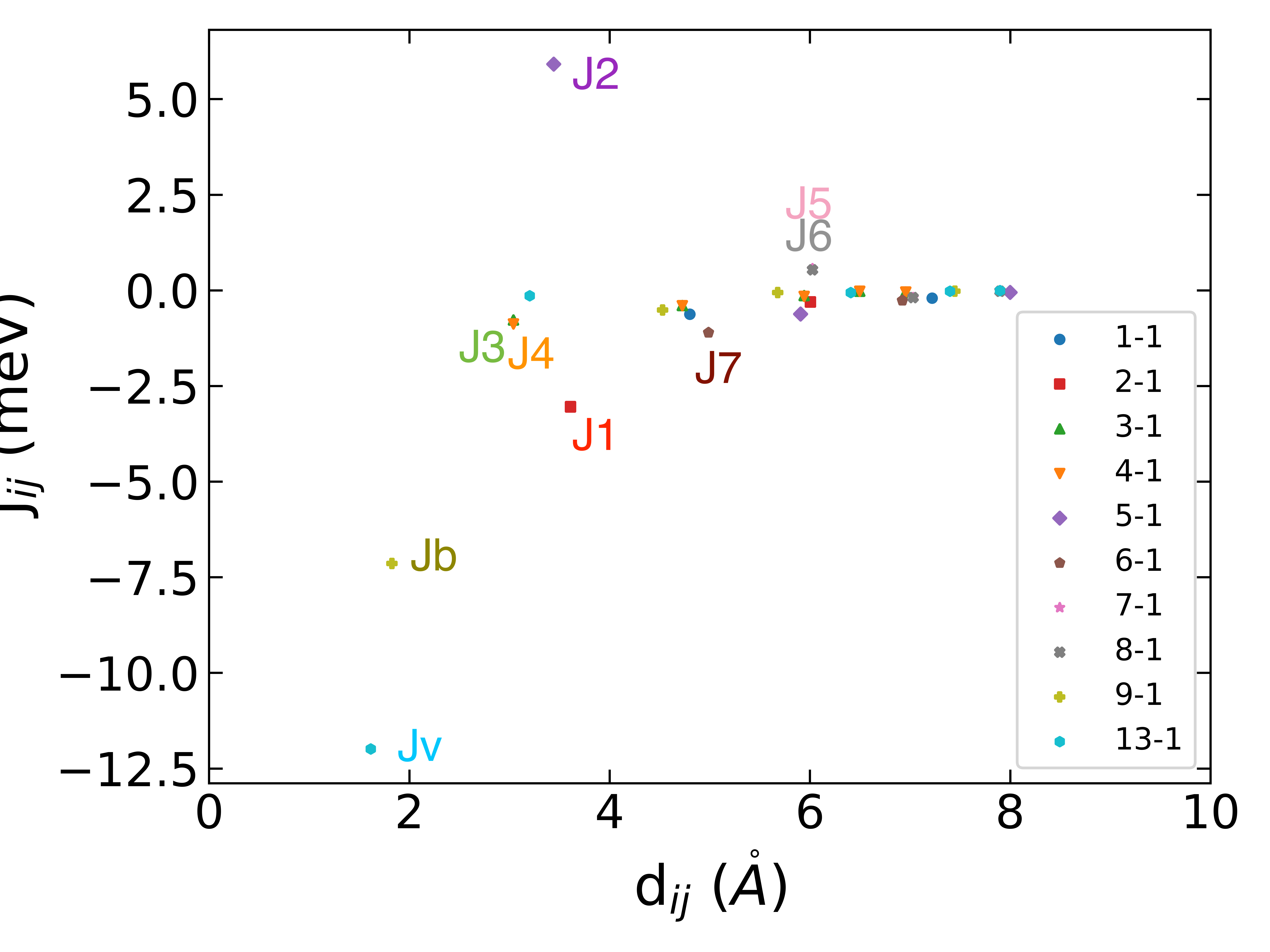}{}
    \caption{}
  \end{subfigure}
  \begin{subfigure}{5cm}\centering
    \includegraphics[width=5cm]{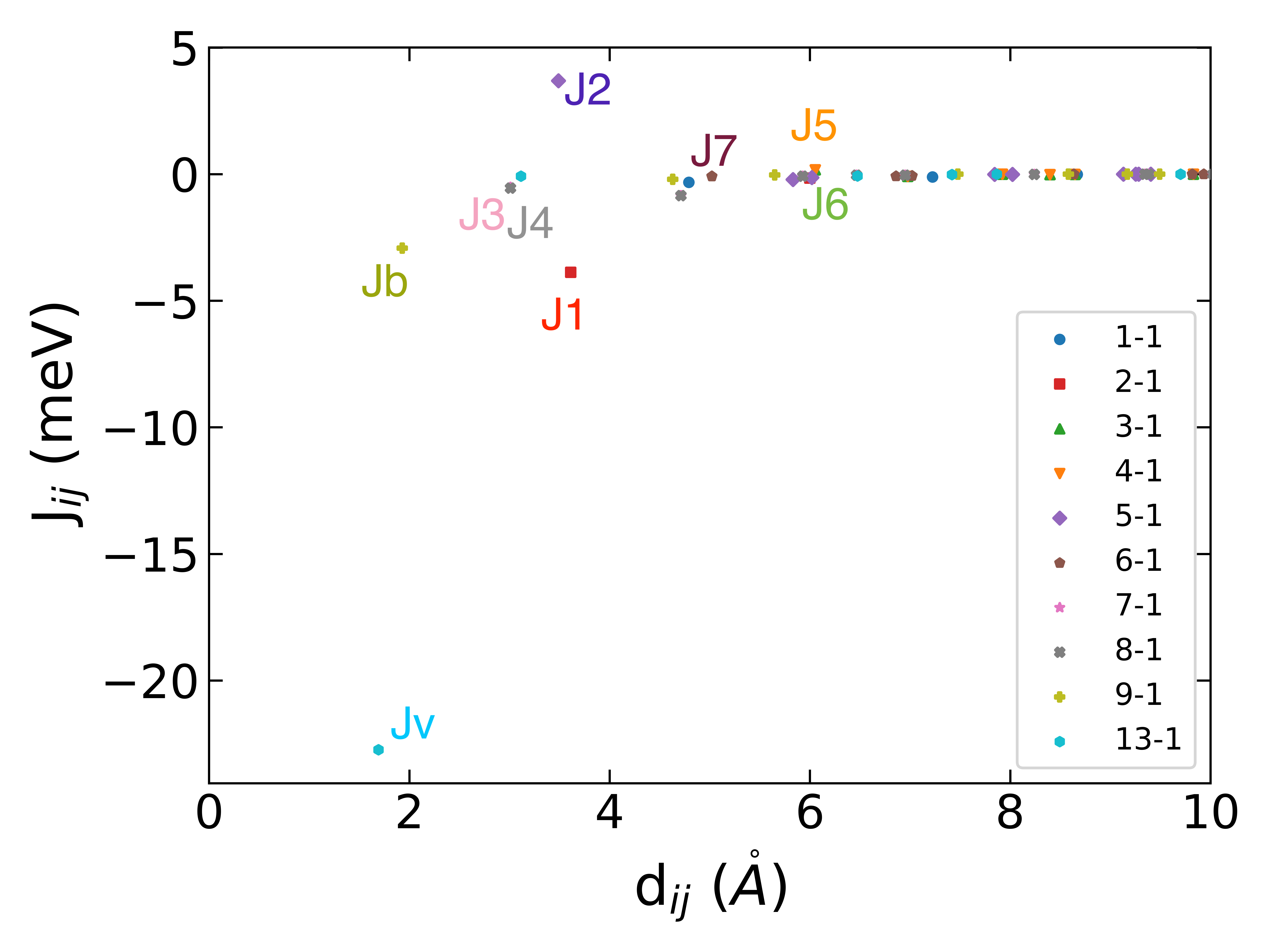}{}
    \caption{}
  \end{subfigure}
  \begin{subfigure}{5cm}\centering
    \includegraphics[width=5cm]{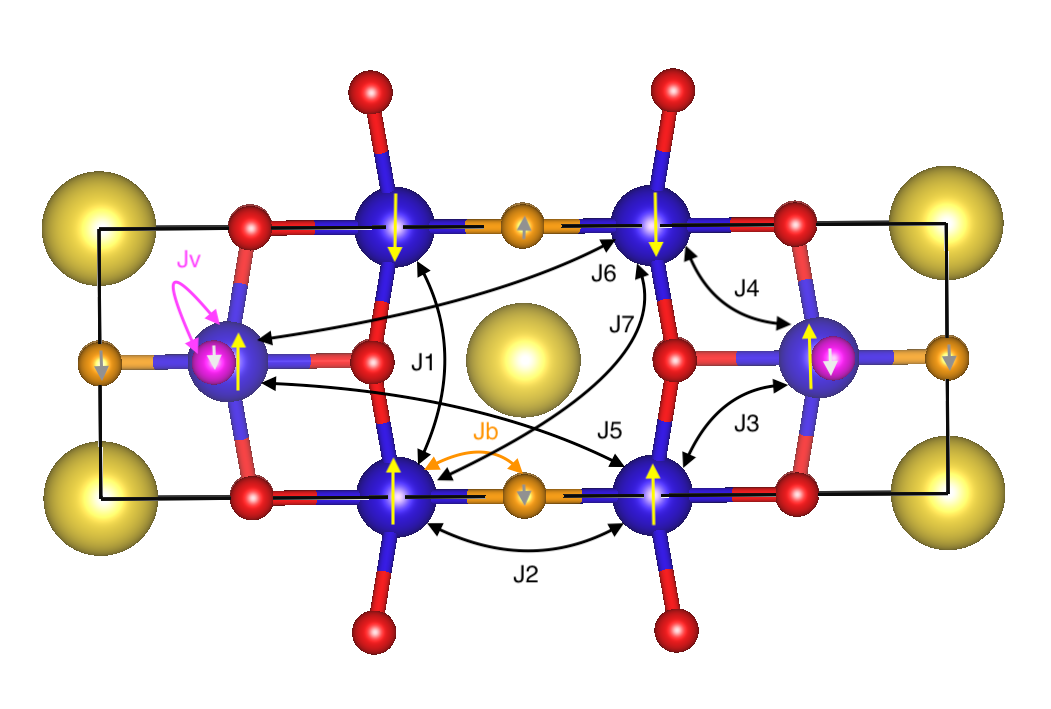}{}
    \caption{}
  \end{subfigure}
      \caption{Exchange interactions in NaV$_2$O$_5$: (a) $Pmmn$, (b) $P2_1mn$, (c) labeling of $J_n$.}
  \label{figexchange-na}
\end{figure*}

\begin{figure*}[hb]
    \includegraphics[width=8cm]{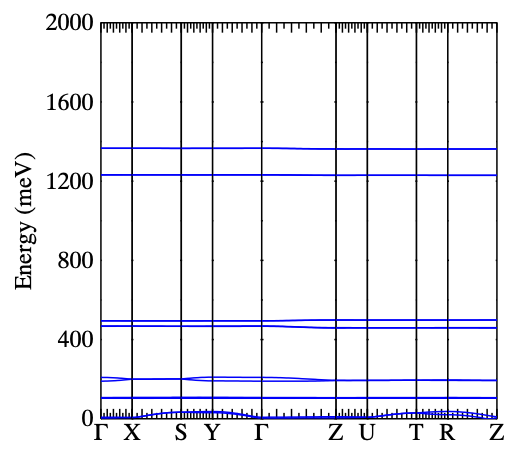}
  \caption{Spin waves dispersion in NaV$_2$O$_5$ $P2_1mn$} \label{figspinwnav2o5}
\end{figure*}

The main exchange interactions in NaV$_2$O$_5$ are given in Table \ref{tabexchange-na} as calculated when including all atoms as magnetic atoms. They differ somewhat from the results of \cite{Lambrechtnav2o5} in which only V, O$_v$ and O$_b$ were treated as magnetic atoms. They are shown as function of distance in Fig.\ref{figexchange-na} for the two structures. They give a $T_N^{MF}=540$ K after the factor $(S+1)/S=3$ correction for $Pmmn$ and $620$ for $P2_1mn$. Overall, we may note that there is a smaller difference between the values for the two structures than in the LiV$_2$O$_5$ case. The spin waves in  $Pmmn$ NaV$_2$O$_5$
were already presented in \cite{Lambrechtnav2o5}.
The spin waves in $P2_1mn$ are given in Fig.\ref{figspinwnav2o5}.

\end{document}